\documentclass[10pt,journal,twocolumn,twoside]{autart} 
\usepackage{graphicx}
\usepackage{amssymb}
\usepackage{amsmath}

\usepackage{algpseudocode}
\usepackage{algorithm}

\usepackage{subcaption}
\usepackage{booktabs}
\usepackage{mathtools}

\usepackage{comment}
\usepackage{color}
\usepackage{cite}

\usepackage{scalerel}
\usepackage{siunitx}
\usepackage{gensymb}

\newcommand{\xd}[0]{\dot{x}}
\newcommand{\zd}[0]{\dot{z}}
\newcommand{\bg}[0]{\bar{g}}

\newcommand{\Aclo}[0]{A_{\mathrm{cl}}}

\newcommand{\Hilbert}[0]{\mathcal{H}}

\newcommand{\xref}[0]{x_{\mathrm{ref}}}

\newcommand{\uref}[0]{u_{\mathrm{ref}}}

\newcommand{\R}[0]{\mathbb{R}}
\newcommand{\Z}[0]{\mathbb{Z}}
\newcommand{\D}[0]{\mathbb{D}}
\newcommand{\N}[0]{\mathbb{N}}
\newcommand{\I}[0]{\mathbb{I}}

\newcommand{\Nr}[0]{\mathbb{N}_{[1,r]}}

\newcommand{\Xterminal}[0]{\mathbb{X}_\mathrm{f}}
\newcommand{\GP}[0]{\mathcal{GP}}

\newcommand{\Vdelta}[0]{V_{\delta}}

\newcommand{\M}[0]{\mathcal{M}}

\newcommand{\norm}[1]{\lVert#1\rVert}

\newcommand{\Mhalf}[0]{M^{\frac{1}{2}}}

\newcommand{\Mhalfarg}[1]{M(#1)^{\frac{1}{2}}}

\newcommand{\Mgammahalf}[1]{M(\gamma^\star(#1))^{\frac{1}{2}}}
\newcommand{\Mgammamhalf}[1]{M(\gamma^\star(#1))^{-\frac{1}{2}}}
\newcommand{\Mgamma}[1]{M(\gamma^\star(#1))}

\newcommand{\gOpt}[0]{\gamma^\star(s)}
\newcommand{\gOpts}[0]{\gamma_s^\star(s)}
\newcommand{\gOptd}[0]{\dot{\gamma}^\star(s)}
\newcommand{\gOptsd}[0]{\dot{\gamma}_s^\star(s)}
\newcommand{\gammaU}[0]{\gamma^u(s)}
\newcommand{\gammaW}[0]{\gamma^w(s)}

\newcommand{\ds}[0]{\text{d}s}
\newcommand{\dstilde}[0]{\text{d}\tilde{s}}
\newcommand{\dtau}[0]{\text{d}\tau}

\newcommand{\traj}[0]{_{\tau|t}}
\newcommand{\tauts}[0]{\tau+T_s}
\newcommand{\tauhatTs}[0]{\hat{\tau}+T_\mathrm{s}}
\newcommand{\tauhat}[0]{\hat{\tau}}

\newcommand{\pder}[2]{\frac{\partial{#1}}{\partial{#2}}}

\newcommand{\RAMPC}{GP-RAMPC}
\newcommand{\RMPC}{GP-RMPC}

\newbool{arxiv} 
\setbool{arxiv}{true}

\usepackage[T1]{fontenc}
\usepackage[utf8]{inputenc}

\newtheorem{theorem}{\bf Theorem}
\newtheorem{assumption}{\bf Assumption}
\newtheorem{proposition}{\bf Proposition}
\newtheorem{lemma}{\bf Lemma}
\newtheorem{remark}{\bf Remark}

\begin{document}

\allowdisplaybreaks

\begin{frontmatter} 
\title{A robust and adaptive MPC formulation for\\ Gaussian process models}
\author[IDSC]{Mathieu Dubied}, 
\author[IDSC]{Amon Lahr}, 
\author[IDSC]{Melanie N. Zeilinger}, 
\author[ICL,IDSC]{Johannes K{\"o}hler}\ead{j.kohler@imperial.ac.uk} 
\address[IDSC]{Institute for Dynamic Systems and Control, ETH Zurich, Zurich, Switzerland}%
\address[ICL]{Department of Mechanical Engineering, Imperial College London, London, UK}
\thanks{This work has been supported by the Swiss National Science Foundation under NCCR Automation (grant agreement 51NF40 180545),
and by the European Union's Horizon 2020 research and innovation programme, Marie~Sk\l{}odowska-Curie grant agreement No. 953348,~\mbox{ELO-X}.}
\date{ \copyright  2026 Elsevier Ltd.  All rights reserved.}
\begin{abstract}
In this paper, we present a robust and adaptive model predictive control (MPC) framework for uncertain nonlinear systems affected by bounded disturbances and unmodeled nonlinearities.
We use Gaussian Processes (GPs) to learn the uncertain dynamics based on noisy measurements, including those collected during system operation. 
As a key contribution, we derive robust predictions for GP models using contraction metrics, which 
are incorporated in the MPC formulation. 
The proposed design guarantees recursive feasibility, robust constraint satisfaction and convergence to a reference state, with high probability. 
We provide a numerical example of a planar quadrotor subject to difficult-to-model ground effects, which highlights significant improvements achieved through the proposed robust prediction method and through online learning.
\end{abstract}
\end{frontmatter} 
 
\section{Introduction} 
Model predictive control (MPC) is an optimization-based control technique that can ensure high-performance control for nonlinear systems, while ensuring satisfaction of safety-critical constraints~\cite{rawlings2017model}. 
However, this guarantee on safe operation typically assumes that an accurate model of the system is available, which is rarely the case in practice.
Given a bound on the model uncertainty, theoretical guarantees on safe system operation can be derived using robust MPC approaches~\cite{houska2019robust}. By further using online data to adapt the model, robust adaptive MPC approaches enhance performance during online operation~\cite{adetola2011robust,kohler2021robust,sasfi2023RAMPC}. 
However, existing methods are largely limited to uncertainty that is linearly parametrized by some uncertain, finite-dimensional parameter vector. 
In contrast, Gaussian processes (GPs) are a modern machine learning technique to learn general unknown functions from data~\cite{rasmussen2006gaussian},
while providing rigorous uncertainty estimates~\cite{chowdhury2017kernelized,fiedler2021practical}.
In this paper, we derive an MPC formulation that uses GPs to model general unknown dynamics, updates these models during operation, and theoretically guarantees safe operation during closed-loop operation.

The development of GP-MPC formulations has seen a lot of progress in recent years, including tailored numerical algorithms ~\cite{polcz2023efficient,lahr2023zero,lahr_l4acados_2024} and experimental applications to different robots~\cite{ostafew2016robust,kabzan2019learning,torrente2021data}. 
Despite the demonstrated empirical success, these methods largely fail to provide theoretical guarantees on constraint satisfaction, see the review paper~\cite{scampicchio2025gaussian} for a detailed discussion. 
Existing GP-MPC approaches entail moment-based approximations~\cite{girard2002gaussian,hewing2019cautious}, sampling-based approximations~\cite{prajapat2025sampling}, multi-step GP models~\cite{maddalena2021kpc,pfefferkorn2022exact}, and robust MPC techniques using sequential propagation~\cite{koller2018learning,cao2025optimistic,polver2025robust}

We are particularly interested in sequential robust propagation techniques, as they can yield strong theoretical guarantees and computationally efficient implementations.
Specifically, they combine high-probability error bounds for GP regression~\cite{chowdhury2017kernelized,fiedler2021practical} with  robust MPC~(RMPC) techniques, which sequentially predict robust reachable sets~\cite{houska2019robust}. 
While such \RMPC{} approaches have been proposed in~\cite{koller2018learning,cao2025optimistic,polver2025robust}, they can be prohibitively conservative: 
\cite{cao2025optimistic}~utilize interval arithmetics to sequentially predict boxes 
and \cite{polver2025robust} predicts balls using Lipschitz continuity of the dynamics. 
These approaches often lead to exponentially growing reachable sets, even for stable linear systems. 
\cite{koller2018learning} sequentially optimizes ellipsoidal reachable sets using linearized dynamics. However, the accumulating linearization error can also lead to an exponential growth of the reachable set; see also the numerical comparison in Section~\ref{sec:numerical_example}. 
Overall, existing \RMPC{} approaches~\cite{koller2018learning,cao2025optimistic,polver2025robust} are often overly conservative, are not compatible with online model updates, or cannot ensure recursive feasibility.  

To ensure recursive feasibility and constraint satisfaction despite model mismatch and online model updates, modern robust~\cite{zhao2022tube,kohler2020computationally,rakovic2022homothetic,sasfi2023RAMPC} and robust adaptive~\cite{lopez2019adaptive,kohler2021robust,sasfi2023RAMPC} MPC approaches have been proposed for parametric uncertainties. 
To model a larger class of uncertain dynamics, it is desirable to expand such approaches to non-parametric models, like GPs.
However, it is not straightforward to extend the required monotonicity properties to non-parametric settings, see discussion in Sections~\ref{sec:RMPC_algorithms_theoretical_properties} and \ref{subsec:RAMPC_theoretical_properties} later for details.
Overall, a computationally efficient GP-MPC that ensures safe operation while learning the dynamics online is currently not available.

\textit{Contribution: }
We propose a \textit{robust} and \textit{adaptive} MPC (RAMPC) formulation for Gaussian process models, i.e., we utilize high-probability bounds of GPs for \textit{robust} predictions and we \textit{adapt} the model during runtime using online collected data. 
The proposed approach
\begin{enumerate}
    \item uses online data to perform GP updates during system operation, thus reducing uncertainty;
    \item is applicable to a large class of nonlinear uncertain continuous-time systems;
    \item comes with closed-loop theoretical guarantees on recursive feasibility, constraint satisfaction, and convergence, with a user-chosen probability (Thm.~\ref{thm:RMPC_feasibility_convergence}/\ref{thm:RAMPC_feasibility_convergence});  
    \item uses a computationally efficient reachability analysis, which augments the prediction model with a scalar dynamics component depending on the GP mean and covariance.
\end{enumerate}
The proposed approach uses an offline-constructed contraction metric to parametrize the reachable set, which builds on existing work on nonlinear RMPC using contraction metrics~\cite{zhao2022tube,sasfi2023RAMPC} and addressing  state-dependent model-mismatch~\cite{kohler2020computationally,sasfi2023RAMPC}. 
In particular, we derive a scalar differential equation that utilizes the state-dependent error bounds from GPs to scale the reachable set around the nominal prediction such that it contains all possible future trajectories (Thm.~\ref{thm:RMPC_tube}), see Section~\ref{sec:RMPC_algorithms_theoretical_properties} regarding novelty compared to existing methods.\\ 
To establish recursive feasibility of the proposed \RAMPC{} scheme,
we employ a collection of GP models, 
deriving a non-increasing tube using set intersection arguments, and use an online-optimized linear combination of the GP posteriors means for the nominal prediction. 
These novel techniques address the problem  that the GP error bounds are not nested and that the most recent posterior mean prediction is in general not a feasible nominal prediction.
The optimization over the nominal model is motivated by existing RAMPC schemes for linearly parametrized models~\cite{sasfi2023RAMPC}, see Section~\ref{subsec:RAMPC_theoretical_properties} for a detailed comparison with related work. \\
Overall, the proposed approach is applicable to nonlinear continuous-time systems with (i)
bounded and sub-Gaussian noise, 
(ii) unmodeled nonlinearities lying in the reproducing kernel Hilbert space (Asm.~\ref{assumption:measurement_noise_sequence_g_RKHS}), 
and (iii) exponentially stabilizable dynamics in terms of contraction metrics (Asm.~\ref{assumption:CCM}). 
\ifbool{arxiv}{

\textit{Outline: }
The problem setup is introduced in Section~\ref{sec:prelim}. 
We first present a \RMPC{} scheme that uses a GP model conditioned on offline data only (Sec.~\ref{sec:RMPC_framework}). 
Then, we incorporate measurements gathered during system operation in a \RAMPC{} scheme to update the model and the uncertainty description online (Sec.~\ref{sec:RAMPC_framework}). 
We provide a numerical example of a planar quadrotor subject to difficult-to-model ground effects, which highlights benefits of the proposed framework (Sec.~\ref{sec:numerical_example}). 
Appendix~\ref{sec:proofs} details the mathematical proofs and Appendix~\ref{sec:vector_valued_g} generalizes the approach to multi-output GPs.}{}

\textit{Notation: }
For a vector~$x\in \R^n$, we express its Euclidean norm by~$\norm{x}=\sqrt{x^\top x}$. For a matrix $M\in\R^{n\times m}$,~$\norm{M}$ denotes the induced 2‐norm, i.e., the maximum singular value of~$M$. Weighted vector norms with respect to a positive definite and symmetric matrix~$M$ are expressed as~\mbox{$\norm{x}_M=\sqrt{x^\top M x}$} and by $\Mhalf$ we denote the Cholesky decomposition, i.e.,~$\left(\Mhalf\right)^\top \Mhalf = M$. For symmetric matrices~$P$ and~$Q$, the notation~\mbox{$P\preceq Q$} indicates that~$Q-P$ is positive semi‐definite. 
The identity matrix of size~$n\times n$ is given by~$I_n$, the non-negative real numbers by~$\R_{\geq 0}$, and the non-negative integers by~$\N_{\geq 0}$. In addition, we denote by~$\Nr$ the set of integers from~$1$ to~$r$.
The cardinality of a set~$\M$ of~$N$ elements is given by~$\#\M=N$. Given a random variable~$X$, its  expected value is denoted by~$\mathbb{E}[X]$ while~$\mathbb{P}[X\in \mathbb{S}]$ denotes the probability that~$X$ takes a value in the set~$\mathbb{S}$. Finally, for a given continuously differentiable function~$f(x,u)$, we define its partial derivative with respect to~$x$, evaluated at a point~$(z,v)$, as~$\pder{f}{x}\big|_{(z,v)}$.

\section{Problem setup \& preliminaries} \label{sec:prelim}
In this section, we introduce the problem setup (Sec.~\ref{subsec:problem_setup}) and provide preliminaries regarding model learning using GPs (Sec.~\ref{subsec:background_GPs}).

\subsection{Problem setup} \label{subsec:problem_setup}
We consider a continuous-time nonlinear system
\begin{equation}\label{eq:system_definition_general}
    \xd(t) = f_w(x(t), u(t), g, d(t)), \quad x(0)=x_0,
\end{equation}
where~\mbox{$x(t)\in\R^n~$} is the state of the system at time~\mbox{$t \geq 0$} with initial state~\mbox{$x(0) = x_0$}, ~\mbox{$u(t)\in \R^m$} the control input,~\mbox{$g:\R^n\rightarrow \R$} an unknown function, and~\mbox{$d(t)\in \D$}, a disturbance lying in the (known) compact set~$\D\subset\R^q$. We assume that~$f_w$ is linear\footnote{%
Linearity in $u,d$ can be achieved using state augmentation, while the unknown function $g$ can always be defined such that the model error is linear in $g$.} in~$u,g,d$, i.e., 
\begin{equation}\label{eq:system_definition}
    f_w(x,u,g,d) = f(x) + B(x)u + Gg(x) + E(x)d,
\end{equation}
\noindent where~$f$ is a known function,~$B(x)\in\R^{n\times m}$ and~$E(x)\in\R^{n\times q}$ are state-dependent matrices, and~$G\in\R^{n}$ is a constant vector. While we consider a scalar function~$g$ for simplicity of exposition, we show in \ifbool{arxiv}{Appendix~\ref{sec:vector_valued_g}}{\cite[Appendix B]{dubied2025robust}} how the derived results generalize to vector-valued functions~$g$. We assume~$u$ and~$d$ to be piecewise continuous, as well as the function~$f_w$ to be continuously differentiable and Lipschitz continuous with respect to~$x$, so that the existence of a unique solution~$x(\cdot)$ to the differential equation~\eqref{eq:system_definition_general} is guaranteed \cite[Thm.~4.16]{slotine1991applied}.

The system should satisfy the (joint-in-time) probabilistic  state and input constraints
\begin{equation}\label{eq:state_input_constraints}
    \mathbb{P}[\left(x(t),u(t)\right)\in \Z, \forall t\geq 0] \geq 1-p,
\end{equation}
where~$p\in(0,1)$ is chosen by the user and~$\Z$ is a compact set described by~$r$ inequalities: 
\begin{equation}\label{eq:def_Z_set}
    \Z := \left\{(x,u)\in \R^{n+m}\;|\;h_j(x,u)\leq 0, \;j\in\Nr\right\}.
\end{equation}
The proposed design will utilize probabilistic GP bounds and hence all results in this paper will hold jointly with this probability $1-p$, unless specified otherwise. We denote by~$\Z_x$ the projection of~$\Z$ on its~$x$-components and assume $h_j$ to be continuously differentiable. 
The control goal is to stabilize a steady-state~$(\xref(g),\uref(g))$, which may depend on the unknown function~$g$ and is assumed to lie in the interior of the constraints, i.e.,
\begin{equation}
\label{eq:reference_state}
\begin{aligned}
    f_w(\xref(g),\uref(g),g,0) &= 0, \\[1ex] 
    (\xref(g),\uref(g)) &\in \mathrm{int}(\mathbb{Z}).
\end{aligned}
\end{equation}

\subsection{Model learning with Gaussian processes} \label{subsec:background_GPs}
We consider the following assumption to facilitate the learning of the unknown function $g(x)$ from measurements. 

\begin{assumption}\label{assumption:measurement_noise_sequence_g_RKHS} The following hold: \\
    (i) Measurements~$y_i\in\mathbb{R}$ of the form 
    \begin{equation}\label{eq:noisy_measurement}
        y_i = g(x_i) + \varepsilon_i,~i\in\N_{\geq 1}
    \end{equation}
    are available, 
    where~$x_i\in \Z_x$ and the noise sequence~$\{\varepsilon_i\}_{i=1}^\infty$ is conditionally~$R$‐sub‐Gaussian for a fixed constant~$R\geq0$. \\
    (ii) The function~$g$ lies in the reproducing kernel Hilbert space (RKHS) defined by a positive-definite kernel~\mbox{$k: \R^n\times \R^n \rightarrow\R$}, i.e.,~$g\in \Hilbert_k$. Its norm is bounded by a known constant~$B_g>0$, i.e., $\norm{g}_{\Hilbert_k}\leq B_g.$\\
    (iii) 
    The derivative of~$g$ is contained in a known compact set~$\mathbb{S}\subseteq\mathbb{R}^{1\times n}$, i.e.,
    $\pder{g}{x}|_{x} \in\mathbb{S}, \forall x\in \Z_x.$
    
\end{assumption}

\begin{remark}
    Condition (i) considers sub-Gaussian noise, which contains a broad class of distributions commonly considered in the literature. In particular, it contains as a special case independent and identically distributed (i.i.d.) noise with a Gaussian distribution and any distribution with bounded support that is zero mean~\cite{vershynin2018high,ao2025stochastic}.  
    Measurements of the form~\eqref{eq:noisy_measurement} can be naturally obtained 
    from noisy derivative measurements~$\dot{x}$
    subject to sub-Gaussian%
    \footnote{%
    Independently generated process noise $d(t)$ is conditionally independent from the state $x(t)$ with the (causal) dynamical system~\eqref{eq:system_definition_general}. 
    } 
    process noise~$d$ and measurement noise 
    if~$G$ is full rank. Explicitly, $y_i=G^\dagger \left(\tilde{\xd}_i-f(x_i)-B(x_i)u_i\right)$, where $G^\dagger$ is the pseudo-inverse of $G$ and $\tilde{\xd}_i$ is the noisy derivative measurement. In this case, $\varepsilon_i=G^\dagger(E(x_i)d_i + \nu_i)$ is conditionally sub-Gaussian if $E(x)$ is uniformly bounded.
    Condition (ii) is a standard regularity condition, which ensures that the kernel $k$ is well-suited to model the unknown function $g$~\cite{chowdhury2017kernelized,fiedler2021practical,abbasi2013online}.
    We note that, while rigorous and non-conservative upper bounds for~$B_g$ are generally challenging to obtain, in practice it can be set by adding a suitable margin to a lower bound estimated from data (cf. \cite{tokmak2023automatic}). 
    Finally, a set~$\mathbb{S}$ satisfying condition (iii) can be derived, given a Lipschitz continuous kernel~$k$ and using condition (ii) (see, e.g.,~\cite[Proof Prop.~4.30]{steinwart2008support}, or~\cite[Proof Lemma 1]{tokmak2023automatic}).
\end{remark}

Given~$N$ noisy observations~$(x_i,y_i)$, $i=1,\dots,N$, and a zero‐mean prior on~$g$, the posterior mean and variance of a GP at a state~$x\in\mathbb{R}^n$ are given by 
\begin{align}\label{eq:GP_posteriors}
    \mu_N(x) 
    &
    = k_N(x)^\top (K_N+\sigma^2I_N)^{-1}y_N,
    \\
    \sigma_N^2(x)
    &
    = k(x,x)-k_N(x)^\top (K_N+\sigma^2I_N)^{-1}k_N(x),
    \nonumber
\end{align}
where~$k_N(x)= [k(x_1,x),...,k(x_N,x)]^\top\in \R^N$, 
$[K_N]_{ij}=k(x_i,x_j)$ is the Gram matrix~\cite[Sec.~2]{rasmussen2006gaussian}, and~$\sigma>0$ is a user-defined regularization parameter.

The following lemma provides a high-probability bound on the function $g(x)$ using the GP.
\begin{lemma}\label{lemma:GP_estimate_uncertainty_bound}      \cite[Thm.~1]{fiedler2021practical}
        Let Assumption~\ref{assumption:measurement_noise_sequence_g_RKHS} hold. 
        It holds that 
    \begin{align}\label{eq:bound_g_mu_general_expression}
        &\mathbb{P}[ |g(x)-\mu_N(x)| \leq w(x),\forall x\in \mathbb{Z}_x, N\in\N_{\geq 0}]
        \geq 1-p, \\
    \label{eq:uncertainty_bound_fiedler}
                &w(x)=\beta_N \sigma_N(x),\\
                &\beta_N = B_g + \dfrac{R}{\sigma} \sqrt{\ln\left(\det\left(\dfrac{\bar{\sigma}^2}{\sigma^2} K_N + \bar{\sigma}^2 I_N\right)\right) - 2 \ln(p)}, \nonumber
    \end{align}
 and~\mbox{$\bar{\sigma}^2 = \max\{1, \sigma^2\}$}.       
\end{lemma}
Lemma~\ref{lemma:GP_estimate_uncertainty_bound} holds for conditionally independent noise sequences, 
which commonly arise in control applications.
Importantly, the bound~\eqref{eq:uncertainty_bound_fiedler} holds jointly for all~$N$, meaning for all GPs conditioned on data gathered successively. 
While the following exposition utilizes the GP error bound from Lemma~\ref{lemma:GP_estimate_uncertainty_bound}, it can naturally accommodate any other bound $w$ satisfying Inequality~\eqref{eq:bound_g_mu_general_expression}. A selection of different bounds~$w$ under different noise assumptions is given in
\ifbool{arxiv}{Appendix~\ref{sec:probability_bounds_table}
}
{~\cite[Appendix~C.1]{dubied2025robust}. 
}
The \RMPC{} scheme proposed in Section~\ref{sec:RMPC_framework} considers the GP based on $N$ offline measurements, while the \RAMPC{} scheme introduced in Section~\ref{sec:RAMPC_framework} additionally uses online measurements to update the GP.

\section{Robust GP-MPC} \label{sec:RMPC_framework}
In this section, we present a \RMPC{} framework that leverages the GP uncertainty bound to make robust predictions of the system's state.  
\ifbool{arxiv}{We first introduce the main conceptual idea in Section~\ref{subsec:nominal_sys_description}.
After presenting the corresponding optimal control problem (OCP) in Section~\ref{subsec:RMPC_ocp}, we elaborate on
 the uncertainty propagation in~Section~\ref{subsec:tube_construction} 
We end this section by summarizing the resulting MPC algorithm and presenting its theoretical properties in Section~\ref{sec:RMPC_algorithms_theoretical_properties}.}{}

\subsection{Robust prediction for uncertain systems}\label{subsec:nominal_sys_description}
To ensure that the system's states satisfy the constraint~\eqref{eq:state_input_constraints} without knowing its exact dynamics, we use a robust approach that constructs a tube around a nominal prediction~$z_t$. This tube is then guaranteed to contain all possible trajectories~$x(t)$ of the true system~\eqref{eq:system_definition_general} with a user-specified probability and hence can be utilized for robust planning. 
The nominal predictions are given by
%
%
%
    \begin{align}
\label{eq:nominal_system}
         \zd_\tau 
        &
        = f_w(z_\tau , v_\tau , \bg, 0) 
        =: \bar{f}(z_\tau ,v_\tau ,\bg),~\tau\geq 0.
    \end{align}
Compared to the real system~\eqref{eq:system_definition_general}, this nominal prediction neglects the disturbances ($d=0$) and uses an estimate~\mbox{$\bg:\R^n\rightarrow \R$} in place of the true (unknown) function~$g$. 
For the \RMPC{} formulation, the function~$\bg$ is given by the posterior mean~$\mu_N$ of the GP trained on~$N$ offline data points~\eqref{subeq:RMPC_formulation_constraint_2.5}.
The tube around this nominal prediction will be of the form $\mathbb{T}_{\tau}=\{x\in\mathbb{R}^n|~V_\delta(x,z_\tau)\leq \delta_\tau\}$, where $V_\delta:\mathbb{R}^n\times\mathbb{R}^n\rightarrow\mathbb{R}_{\geq 0}$ is an appropriate distance function that is designed offline and $\delta_\tau\geq 0$ is a scaling that is predicted online. 
Since our goal is to stabilize the reference~\eqref{eq:reference_state}, we consider a standard quadratic stage cost
\begin{equation}\label{eq:def_quadratic_stage_cost}
    \ell(z,v,\bg)= \norm{z-\xref(\bg)}_{Q_c}^2 + \norm{v-\uref(\bg)}_{R_c}^2,
\end{equation}
where~$Q_c\in\R^{n\times n}$ and~$R_c\in \R^{m\times m}$ are positive-definite matrices and $\bg$ is the estimate of the unknown function.

\subsection{\RMPC{} optimal control problem}\label{subsec:RMPC_ocp}
The proposed \RMPC{} scheme is characterized by the following OCP:
\begin{subequations}\label{eq:RMPC_optimization_problem}
    \begin{align}
        \label{subeq:RMPC_cost_function}
        \min_{v_{\cdot|t},z_{\cdot|t},\delta_{\cdot|t}}& \; \int_{0}^{T_\mathrm{f}}\ell(z\traj,v\traj,\bg\traj) \dtau+
         \ell_{\mathrm{f}}(z_{T_\mathrm{f}|t},\bg\traj)
        \\
        \label{subeq:RMPC_formulation_constraint_1}
        \text{s.t. }\quad 
        &\dot{z}\traj=\bar{f}(z\traj,v\traj,\bg\traj),
        \\
        \label{subeq:RMPC_formulation_constraint_2}
        &\dot{\delta}\traj=f_{\delta}(z\traj,\delta\traj,w\traj),
        \\   
         \label{subeq:RMPC_formulation_constraint_2.5}
         &\bg\traj(\cdot) = \mu_N(\cdot),
         \\
         \label{subeq:RMPC_formulation_constraint_2.75}
         & w\traj(\cdot) = \beta_N\sigma_N(\cdot),
        \\   
        \label{subeq:RMPC_formulation_constraint_3}
        &h_j(z\traj,v\traj)+c_j\delta\traj\leq 0,
        \\
        \label{subeq:RMPC_formulation_constraint_4}
        &(z_{T_\mathrm{f}|t},\delta_{T_\mathrm{f}|t},\bg_{T_\mathrm{f}|t}, w_{T_\mathrm{f}|t})\in \Xterminal,
        \\   
        \label{subeq:RMPC_formulation_constraint_5}
        &V_\delta(x(t),z_{0|t}) \leq \delta_{0|t},  
        \\
        \label{subeq:RMPC_formulation_constraint_6}
        & \tau\in[0,T_\mathrm{f}],\; j\in\N_{[1,r]},
    \end{align}
\end{subequations}
where the notation~$\tau|t$ denotes a prediction of a quantity for the future time~$t+\tau$ computed at the current time~$t$.
Problem~\eqref{eq:RMPC_optimization_problem} is a natural generalization of existing homothetic tube MPC formulations, such as~\cite{kohler2020computationally,rakovic2022homothetic,sasfi2023RAMPC}. In particular, the cost~\eqref{subeq:RMPC_cost_function} depends on a nominal state and input trajectory $z,v$, which are predicted using nominal dynamics~\eqref{subeq:RMPC_formulation_constraint_1}. 
In addition, a tube scaling $\delta\geq 0$ is predicted~\eqref{subeq:RMPC_formulation_constraint_2} and used to tighten the state and input constraints~\eqref{subeq:RMPC_formulation_constraint_3}. 
Notably, the nominal dynamics and the dynamics of this tube scaling depend on the mean and uncertainty bound on the GP using~\eqref{subeq:RMPC_formulation_constraint_2.5}. 
The initial condition~\eqref{subeq:RMPC_formulation_constraint_5} links the (optimized) initial state of the nominal trajectory to the current real state~$x(t)$, as well as the tube scaling~$\delta_{0|t}$. Lastly, a general terminal set constraint is imposed~\eqref{subeq:RMPC_formulation_constraint_4}, together with the terminal cost $\ell_f$, whose properties are discussed in Section~\ref{sec:RMPC_algorithms_theoretical_properties}. 
The MPC problem~\eqref{eq:RMPC_optimization_problem} is solved at each discrete sampling time $t_k=T_{\mathrm{s}} k$, $k\in\N$ and the closed-loop input is given by
\begin{equation}\label{eq:RMPC_control_input_from_ocp}
    u(t)=\kappa(x(t),z^\star_{\cdot|t_k}, v^\star_{\cdot|t_k}),
\end{equation}
where~$z^\star_{\cdot|t_k}$ and~$v^\star_{\cdot|t_k}$ are the optimal\footnote{%
We assume that a minimizer to Problems~\eqref{eq:RMPC_optimization_problem} and \eqref{eq:RAMPC_optimization_problem} exists, see~\cite[Prop.~2]{fontes2001general} for sufficient conditions.} nominal state and input over the sampling period~$[t_k,t_k+T_\mathrm{s})$ and $\kappa$ is a later-specified feedback law. 
For computational tractability, we parametrize the nominal input $v_{\tau|t}$ as piece-wise constant across each sampling interval $\tau\in [i T_{\mathrm{s}},iT_{\mathrm{s}}+T_{\mathrm{s}})$, $i\in\N_{[0,N_{\mathrm{f}}-1]}$, with $N_{\mathrm{f}}:=T_{\mathrm{f}}/T_{\mathrm{s}} \in\N$.
In the following, we detail the ingredients of the OCP~\eqref{eq:RMPC_optimization_problem}.

\subsection{Tube construction using contraction metrics}\label{subsec:tube_construction}
We construct the tube using an (incremental) Lyapunov function $V_\delta(x,z)$ and a stabilizing controller $\kappa(x,z,v)$. These ingredients are derived using contraction metrics, which are constructed offline to satisfy the following conditions.  
\begin{assumption}\label{assumption:CCM}
    There exists a continuously differentiable, symmetric matrix~$M(x): \R^n\rightarrow \R^{n\times n}$, a matrix~$K(x): \R^n \rightarrow\R^{m\times n}$, a contraction rate~$\rho> 0$, and positive definite matrices~$\underline{M}\in \R^{n\times n}$ and~$\overline{M}\in \R^{n\times n}$, such that for all~\mbox{$(x,u)\in \Z$}, $g_x\in\mathbb{S}$, \mbox{$d\in \D$}:%
    \begin{subequations}
      \label{subeq:M}
        \begin{align}
            \label{subeq:M_Acl_bound}
            &
            \dot{M}(x)  
            +  \Aclo(x,u,g_x,d)^\top M(x)
            \nonumber
            \\
            &
            + M(x) \Aclo(x,u,g_x,d)\preceq -2\rho M(x),
            \\
            &\underline{M} \preceq M(x) \preceq \overline{M},\label{subeq:M_bound}
        \end{align}
    \end{subequations}
    with~$\xd=f_w(x,u,g,d)$, and
    \begin{align}\label{eq:Aclo_definition}
        \Aclo (x,u,g_x,d)
        :=&\pder{f_w}{x}\bigg|_{(x,u,0,d)} + \pder{f_w}{u}\bigg|_{(x,u,0,d)}K(x) 
        \nonumber
        \\
        &
        + G g_x.
    \end{align}
\end{assumption}
Conditions~\eqref{subeq:M} can be reformulated as linear matrix inequalities (LMIs) using standard re-parametrizations~\cite{zhao2022tube,manchester2017control}. 
Note that~\eqref{eq:Aclo_definition} does not depend on the unknown function $g$, but only the known bounds on the Jacobian $g_x\in\mathbb{S}$ (Asm.~\ref{assumption:measurement_noise_sequence_g_RKHS}).  
Furthermore, $\dot{M}$ can be made independent of the unknown function $g$ 
by parametrizing $M(x)$ such that $\frac{\partial M(x)G}{\partial x}\equiv 0$~\cite[Sec.~3.A]{manchester2017control}. 

The contraction metric conditions~\eqref{subeq:M} are sufficient to guarantee exponential incremental stability with a suitable feedback controller~\cite{manchester2017control}. 
These conditions are also necessary under mild assumptions~\cite[Thm.~2.1]{tsukamoto2021contraction}.
Nonetheless, the resulting LMIs may be infeasible for some nonlinear systems and parametrizations of $M(x),K(x)$, in which case the proposed approach cannot be applied. 
A comprehensive overview of nonlinear control using contraction metrics is provided in~\cite{tsukamoto2021contraction}. 

%
The corresponding Lyapunov function is given by the Riemannian distance induced by the metric $M(x)$:
\begin{equation}\label{eq:def_Lyapunov_function}
    \Vdelta(x,z) := \min_{\gamma\in \Gamma(z,x)}\int_0^1\|\gamma_s(s)\|_{M(\gamma(s))} \ds.
\end{equation}
Here, $\Gamma(z,x)$ denotes the set of piece-wise smooth curves
\begin{equation}
    \gamma:[0,1]\rightarrow \R^n,\quad \text{with}\quad \gamma(0)=z,\gamma(1)=x,
\end{equation}
with the derivative $\gamma_s(s):=\pder{\gamma(s)}{s}|_s.$
A minimizer to~\eqref{eq:def_Lyapunov_function}, denoted by~$\gamma^\star$, is called \emph{geodesic}, and existence follows from the uniform bounds~\eqref{subeq:M_bound}, see~\cite[Lemma~1]{manchester2017control}. 
The feedback associated with contraction metrics (cf.~\cite{manchester2017control, zhao2022tube}) is given by:
\begin{align}\label{eq:def_feedback_input}
    \kappa(x,z,v) := &\gamma^u(1),\\
\label{eq:def_feedback_input_part_2}
    \gamma^u(s) :=& v + \int_0^sK(\gamma^\star(\tilde{s}))\gamma_s^\star(\tilde{s})\dstilde,
\end{align}
where $\gamma^\star\in\Gamma(z,x)$ is the geodesic. 
Given the incremental Lyapunov function $V_\delta$, we parametrize a homothetic tube by considering sublevel sets centered around the nominal trajectory $z$ with a variable scaling $\delta\geq0$. 
The following theorem establishes the dynamics of this scaling $\delta$,
providing an outer-approximation of the reachable set and ensuring constraint satisfaction.
\begin{theorem}\label{thm:RMPC_tube}
    Suppose Assumptions~\ref{assumption:measurement_noise_sequence_g_RKHS} and~\ref{assumption:CCM} hold. Consider an initial state~$x_0\in\R^n$ and trajectories~$z_t,v_t,\delta_t$ that satisfy, for all~$t\geq 0$,
    \begin{subequations}
    \label{eq:RMPC_assumptions}
        \begin{align}
            \label{subeq:RMPC_assumption_1}
            &
            h_j(z_t,v_t)+c_j\delta_t
            \leq 0, \quad j\in\Nr,
            \\
            \label{subeq:RMPC_assumption_2}
            &
            \dot{z}_t
            = \bar{f}(z_t,v_t,\bg),
            \\
            \label{subeq:RMPC_assumption_3}
            &
            \Vdelta(x_0,z_0)
            \leq \delta_0,
            \\
            \label{subeq:RMPC_assumption_4}
            &
            \dot{\delta}_t
            = -\left(\rho - L_G \right)\delta_t
            + G_M w(z_t) +  E_M
            \\  
            \label{subeq:RMPC_assumption_4_part2}
            &
            \phantom{\dot{\delta}_t}=: f_\delta(z_t,\delta_t,w),
        \end{align} 
    \end{subequations}
    where
    \begin{equation*}
        c_j := \max_{(z,v)\in \Z }\left\Vert\left(\pder{h_j}{x}\bigg|_{(z,v)}+\pder{h_j}{u}\bigg|_{(z,v)} K(z)\right)M(z)^{-\frac{1}{2}}\right\Vert,
    \end{equation*}
    $w$ is introduced~\eqref{eq:uncertainty_bound_fiedler}, and~$L_G$,~$G_M$, and~$E_M$ given by
\begin{subequations}\label{eq:def_offline_constants}
        \begin{align}
        \label{eq:def_L_G}
            L_G
            &:=\max_{x,x'\in \Z_x}\max_{g_x\in\mathbb{S}}\left\{\left\Vert M(x)^{\frac{1}{2}}Gg_x M(x')^{-\frac{1}{2}}\right\Vert\right\}, \\
            \label{subeq:def_G_M}
            G_M
            & :=
            \max_{x\in \Z_x}\left\{ \norm{G}_{M(x)} \right\},
            \\
            \label{subeq:def_E_M}
            E_M 
            & := 
            \max_{x\in \Z_x,\, d\in \D}\left\{\norm{E(x)d}_{M(x)}\right\}.
        \end{align}
    \end{subequations}
     Then, the trajectory
    \begin{subequations}
        \begin{align}
            \dot{x}(t)
            &
            =f_w(x(t),\kappa(x(t),z_t,v_t),g,d(t)),
            \\
            x(0)
            &
            =x_0,
        \end{align}
    \end{subequations}
    satisfies
    \begin{subequations}
    \label{eq:RMPC_result}
        \begin{align}
            \label{subeq:RMPC_result_1}
            &
            \mathbb{P}[\Vdelta(x(t),z_t) \leq \delta_t, \forall t\geq 0] \geq 1-p,
            \\
            \label{subeq:RMPC_result_2}
            &
             \mathbb{P}[(x(t),\kappa(x(t),z_t,v_t))\in \Z, \forall t\geq 0] \geq 1-p.
        \end{align}
    \end{subequations}
\end{theorem}
Equation~\eqref{subeq:RMPC_result_1} ensures that we can predict a homothetic tube that contains the true state trajectory with user-defined probability $1-p$.
Furthermore, \eqref{subeq:RMPC_result_2} yields the desired constraint satisfaction condition~\eqref{eq:state_input_constraints}. The proof of Theorem.~\ref{thm:RMPC_tube} and all other theoretical claims are collected in\ifbool{arxiv}{ Appendix~\ref{sec:proofs}.}{~\cite{dubied2025robust}. Similar to~\cite{sasfi2023RAMPC}, the tube dynamics are obtained by bounding the derivative of the incremental Lyapunov function~\eqref{eq:def_Lyapunov_function} using the contraction metrics (Asm.~\ref{assumption:CCM}). Specifically, we bound the difference between the unknown function~$g$ and its approximation~$\bar{g}$ along the geodesic~$\gamma^\star$ using the Lipschitz-like constant~\eqref{eq:def_L_G} and the error bound~\eqref{eq:bound_g_mu_general_expression}}.

 
\subsection{\RMPC{} algorithms and theoretical properties}\label{sec:RMPC_algorithms_theoretical_properties}

Algorithm~\ref{alg:offline_design_RMPC} summarizes the offline computation and Algorithm~\ref{alg:online_design_RMPC} shows the online control strategy of the \RMPC{} scheme, based on the OCP~\eqref{eq:RMPC_optimization_problem}.
\begin{algorithm}[ht]\caption{Offline computations.}\label{alg:offline_design_RMPC}
 Inputs: Model $f_w$~\eqref{eq:system_definition}, sets~$\Z$,~$\D$,~$\mathbb{S}$, constant~$\beta_N$ (Lem.~\ref{lemma:GP_estimate_uncertainty_bound}).
 \begin{algorithmic}[1]
    \State
    Compute contraction metrics: $M(x)$, $K(x)$, $\rho$ (Asm.~\ref{assumption:CCM}). 
    \State Compute constants $L_G$, $G_M$, $E_M$, and $c_j$ for tube propagation \& constraints (Thm.~\ref{thm:RMPC_tube}). 
\end{algorithmic}
\end{algorithm}
\begin{algorithm}[ht]\caption{Online computations (\RMPC{}).}\label{alg:online_design_RMPC}
\begin{algorithmic}[1]
    \For{\textnormal{each sampling time} $t_k=kT_s, k\in \mathbb{I}_{\geq 0}$}
    \State
    Solve the OCP~\eqref{eq:RMPC_optimization_problem}.
    \State
    Apply the feedback~\eqref{eq:def_feedback_input} for $t\in [t_k,t_{k+1})$.
    \EndFor
\end{algorithmic}
\end{algorithm}

Following standard MPC designs~\cite{rawlings2017model}, we consider the following properties of the terminal set~$\Xterminal$.
\begin{assumption}\label{assumption:terminal_set}  
   For any~\mbox{$(z,\delta,\bg,w)\in \Xterminal$}, there exists an input~\mbox{$v_\mathrm{f}\in\R^m$}, such that the trajectories~$\dot{z}_t = \bar{f}(z_\tau,v_\mathrm{f},\bg)$,~$\dot{\delta}_\tau = f_\delta(z_\tau,\delta_\tau,w)$, $t\in[0,T_{\mathrm{s}}]$, with~$\delta_0=\delta$,~$z_0=z$ satisfy
    \begin{enumerate}
    \begin{subequations}
        \item positive invariance:
        \begin{equation}\label{eq:terminal_set_PI}
            (z_{T_\mathrm{s}},\delta_{T_\mathrm{s}},\bg,w)\in\Xterminal;
        \end{equation}
        \item constraint satisfaction:
        \begin{equation}\label{eq:terminal_set_constraint_satisfaction}
            h_j(z_\tau,v_\mathrm{f})+c_j\delta_\tau \leq 0, \;j\in \Nr,\; \tau\in[0,T_{\mathrm{s}}]; 
        \end{equation}
        \item local control Lyapunov function decrease: 
        \begin{align}
        \label{eq:terminal_set_local_CLF}
            \int_0^{T_\mathrm{s}}\ell(z_\tau,v_\mathrm{f},\bg) \dtau
            \leq \ell_\mathrm{f}(z_0,\bg)-\ell_\mathrm{f}(z_{T_\mathrm{s}},\bg);
        \end{align}
        \item monotonicity, for any~ $\hat{w}:\R^n\rightarrow \R_{\geq 0}$, $w(\cdot)\geq\hat{w}(\cdot)$, and $\hat{\delta}\in[0,\delta]$: 
        \begin{align}
        \label{eq:terminal_set_monotonicity}
            &(z,\delta,\bg,w)\in \Xterminal 
            \Rightarrow (z,\hat{\delta},\bg,\hat{w}) \in \Xterminal.
        \end{align}
    \end{subequations}
    \end{enumerate}

\end{assumption}

A simple constructive design satisfying Assumption~\ref{assumption:terminal_set} is given by\ifbool{arxiv}{~Proposition~\ref{prop:terminal_set} in Appendix~\ref{sec:proof_terminal_set}}{~\cite[Proposition~2]{dubied2025robust}}.

 The following theorem summarizes the theoretical properties of the proposed \RMPC{} scheme.
%
%
\begin{theorem}\label{thm:RMPC_feasibility_convergence}
    Let Assumptions~\ref{assumption:measurement_noise_sequence_g_RKHS},~\ref{assumption:CCM}, and~\ref{assumption:terminal_set} hold. Suppose that the OCP~\eqref{eq:RMPC_optimization_problem} is feasible at~$t_0=0$ for initial state~$x_0$. 
    Then, with at least probability $1-p$, the closed-loop system~\eqref{eq:system_definition} with input~\eqref{eq:RMPC_control_input_from_ocp} resulting from Algorithm~\ref{alg:online_design_RMPC} ensures:
    \begin{enumerate}
        \item Recursive feasibility: Problem~\eqref{eq:RMPC_optimization_problem} is feasible for all times $t_k$, $k\in \N_{\geq 0}$;
        \item Closed-loop constraint satisfaction, i.e., \eqref{eq:state_input_constraints} holds;
        \item Convergence: The nominal trajectories converge to the reference state and input, i.e., for~$~\tau\in[0,T_\mathrm{s})$,
    \end{enumerate}
    \begin{equation}~\label{eq:convergence_RMPC}
        \lim_{k\rightarrow \infty}\norm{(z^\star_{\tau|t_k},v^\star_{\tau|t_k})-(\xref(\mu_N),\uref(\mu_N))}=0.
    \end{equation}
\end{theorem}
Theorem~\ref{thm:RMPC_feasibility_convergence} shows three important properties of the proposed \RMPC{} scheme with a user-specified probability: (i) recursive feasibility, (ii)  constraint satisfaction, (iii) convergence of the nominal state to the reference steady-state~\eqref{eq:convergence_RMPC} and, hence, convergence of the real state to a neighborhood bounded by the scaling~$\delta$.

The derived \RMPC{} OCP~\eqref{eq:RMPC_optimization_problem} is closely related to existing nonlinear robust MPC approaches:
The tube is constructed using offline computed contraction metrics~\cite{zhao2022tube,sasfi2023RAMPC} and the online tube propagation reduces to predicting a scaling~$\delta$~\cite{kohler2020computationally,kohler2021robust,lopez2019adaptive,rakovic2022homothetic,sasfi2023RAMPC}. 
The key novelty lies in the treatment of the general uncertain \emph{function} $g(x)$ using GPs, while existing approaches are restricted to \emph{finite-dimensional} uncertain parameters or disturbances. 
A key theoretical contribution is the fact that the proposed design of the tube (Asm.~\ref{assumption:CCM}) can be executed offline, while enabling robust predictions with the online learned (a-priori unknown) GP models (cf. Sec.~\ref{sec:RAMPC_framework}). 
Compared to the GP-MPC in~\cite{koller2018learning}, the proposed approach requires an additional offline design of a contraction metric. Key benefits of the proposed approach are:
(i)~reduced computational complexity through optimizing a scaling $\delta_\tau\in\mathbb{R}$ instead of matrices $P_\tau\in\mathbb{R}^{n\times n}$;
 (ii)~circumventing the accumulating error from the Taylor approximation, resulting in significantly reduced conservatism (cf. Sec.~\ref{sec:numerical_example});
 (iii)~guaranteed recursive feasibility through a simple to design terminal set \ifbool{arxiv}{(Prop.~\ref{prop:terminal_set} in App.~\ref{sec:proof_terminal_set})}{}.
Next, we show how to incorporate online model updates.

\section{Robust Adaptive GP-MPC} \label{sec:RAMPC_framework}

Leveraging the measurements obtained during system operation, we aim at improving our system's model by using successively updated GP models~\eqref{eq:GP_posteriors} of the unknown function~$g$. The \RAMPC{} scheme should update  its  model and the corresponding uncertainty bound such that
    i)~the predicted tube becomes less conservative through the online collected data, 
    ii)~the MPC problem remains recursively feasible.
Achieving both goals is, however, not straightforward. In particular, a successive GP model has not only a reduced uncertainty bound~$w$~\eqref{eq:uncertainty_bound_fiedler}, but also a different posterior mean~$\mu$~\eqref{eq:GP_posteriors}. Therefore, the bounds on the uncertainty function $g$ constructed in~\eqref{eq:bound_g_mu_general_expression} with two successive GP models are in general not contained in each other, which may lead to feasibility issues during closed-loop operation. 
We address this issue by working with a collection of GP models that ensure the desired nestedness.

\ifbool{arxiv}{First, we describe how online updates of GPs are incorporated in the dynamics in Section~\ref{subsec:online_measurements_RAMPC}. These dynamics are then used to formulate the OCP in Section~\ref{subsec:ocp_RAMPC}. Finally, we show that the resulting \RAMPC{} scheme inherits the theoretical properties of the \RMPC{} scheme.}{}

\subsection{Online measurements and model updates}\label{subsec:online_measurements_RAMPC}

The strategy used by the \RAMPC{} scheme consists in creating successive GP models based on successive online measurements.
We denote the initial data set consisting of $N$ offline measurements considered for the \RMPC{} scheme
by $\mathcal{D}_{t_0}=\{(x_i,y_i),~i\in\N_{[1,N]}\}$, with $y_i,x_i$ according to~\eqref{eq:noisy_measurement}. 
During runtime, we recursively expand the data set with  $\mathcal{D}_{t_{k}}=\mathcal{D}_{t_{k-1}}\cup \{(x_{N+k},y_{N+k})\}$, $k\in\N_{\geq 1}$ using online collected data, with $x_{N+k}=x(t_{k})$. 
We denote the posterior distribution of the GP conditioned on the data set~$\mathcal{D}_{t_k}$ at sampling time~$t_k$ by
\begin{equation}\label{eq:def_GP_tk}
    \GP_{k} := \GP(\mathcal{D}_{t_k}).
\end{equation}
%
%
Once a new measurement is available at sampling time~$t_k$, a new GP model is obtained and added to a GP model collection~$\M_{k}$:
\begin{equation}\label{eq:def_M_tk}
    \M_{k} := \M_{{k-1}}\cup \GP_{k},
\end{equation}
with the initial GP model collection~$\M_{0}$ consisting of the initial GP model~$\GP_{0}$ conditioned on offline measurements~$\mathcal{D}_{t_{0}}$. The GP model collection~$\M_{k}$ therefore stores the different GP models created successively.
By using new GPs conditioned on more data, the \RAMPC{} scheme aims at improving the estimation~$\bg$ of the unknown function~$g$. At sampling time~$t_k$, we define a selection set~$\I_k$
containing the set of indices corresponding to the selected GPs from~$\M_k$. The posterior means of the selected GPs are then linearly combined to express the nominal estimate~$\bg$ at time $t$, i.e.,
\begin{equation}\label{eq:bar_g_as_weighted_sum}
    \bg_t(\cdot) = \sum_{i\in \I_k} \lambda_{t,i} \mu_{\GP_i}(\cdot)=:\tilde{g}(\cdot,\I_{k},\lambda_t) ,
\end{equation}
where~$\mu_{\GP_i}$ corresponds to the posterior mean of~$\GP_i$, and~$\lambda_t\in \R^{\#\M_k}$ describes the vector composed of the scaling factors~$\lambda_{t,i}\in\mathbb{R}$, $i\in\I_k$. The estimate~\eqref{subeq:RMPC_formulation_constraint_2.5} used by the \RMPC{} scheme is therefore a special case where~$\M_{k}=\M_{0}$ contains a single GP trained on offline data at all sampling times~$t_k$ and~$\lambda_{0,t}=1$ for all~$t$.
The following proposition shows how we obtain consistent predictions while using the online measurements to update the GP models.  
\begin{proposition}\label{prop:bound_g_g_bar}
    Let  Assumption~\ref{assumption:measurement_noise_sequence_g_RKHS} hold. Consider two GP model collections~$\M_{k}$ and~$\M_{k+1}$ as described by~\eqref{eq:def_M_tk}, GP selections~$\I_{k}$,~$\I_{k+1}$ with~\mbox{$\I_{k}\subseteq \I_{k+1}$}, ~$k\in\N_{\geq 0}$. 
    Then, for all~$z\in\Z_x$,~\mbox{$\lambda\in\R^{\#\M_k}$}, the following properties hold:

    \begin{enumerate}
        \item Uncertainty bound:
    \begin{align}\label{eq:prop_bound_g_g_bar}
        &\mathbb{P}[ |g(z)-\bg(z)| \leq \tilde{w}(z,\bg,\I_{k}),
        \forall x\in \mathbb{Z}_x, k\in\N_{\geq 0}]
        \nonumber
        \\
        &\geq 1-p,
    \end{align}
    where  $\bg:\mathbb{R}^n\rightarrow\mathbb{R}$ is an arbitrary function and 
        \begin{align}\label{eq:def_w_tilde_RAMPC}
            &\tilde{w}(z,\bg,\I_{k}):=
            \\
            &\max\left\{\min_{i\in \I_k}\{\mu_{\GP_i}(z) + \beta_{\GP_i}\sigma_{\GP_i}(z)\} - \bg(z)\right., 
            \nonumber
            \\
            &
            \phantom{:=\max\left\{\right.}\left.
            \bg(z)- \max_{i\in\I_k}\{\mu_{\GP_i}(z) - \beta_{\GP_i}\sigma_{\GP_i}(z)\} \right\};
            \nonumber
        \end{align}   

    \item Monotonicity:
    \begin{align}\label{eq:set_inclusion_w}
        \tilde{w}(z,\bg,\I_{{k+1}})
        \leq  \tilde{w}(z,\bg,\I_{k});
    \end{align}
    \item Consistency: $\exists \tilde{\lambda}\in\R^{\#\M_{k+1}}$, such that
    \begin{equation}\label{eq:g_bar_k_plus_one_g_bar}
        \tilde{g}(z,\I_{k+1},\tilde{\lambda})= \tilde{g}(z,\I_{k},\lambda).
    \end{equation}
    
    \end{enumerate}
    
\end{proposition}

\begin{remark}
    Proposition~\ref{prop:bound_g_g_bar} constructs the uncertainty bound~$w$ by performing a set-intersection of the different GP confidence bounds. Similar intersections are used in set-membership estimation for uncertain parameters~\cite{kohler2021robust} and for confidence bounds for safe exploration with GPs~\cite{prajapat2025safe}. Using Lemma~\ref{lemma:GP_estimate_uncertainty_bound}, these set-intersections contain $g$ and are thus non-empty, at least with probability~$1-p$. 
\end{remark}
By using multiple GP models, Proposition~\ref{prop:bound_g_g_bar} ensures that the uncertainty bound $w$ is non-increasing with new measurements~\eqref{eq:set_inclusion_w} and that we can obtain consistent predictions~\eqref{eq:g_bar_k_plus_one_g_bar}, which is crucial for the theoretical analysis later. 
The following theorem shows that the robust propagation and constraint satisfaction established in Theorem~\ref{thm:RMPC_tube} equally hold with the adapted GP models.
\begin{theorem}\label{thm:RAMPC_tube}
    Suppose Assumptions~\ref{assumption:measurement_noise_sequence_g_RKHS} and~\ref{assumption:CCM} hold and consider a GP model collection~$\M_{k}$. Then, 
    the high-probability bounds~\eqref{eq:RMPC_result} from Theorem~\ref{thm:RMPC_tube} also hold when using the uncertainty bound $w$ according to~\eqref{eq:def_w_tilde_RAMPC}, any $\lambda\in\R^{\#\M_{k}} $, and any (Lipschitz-continuous) function $\bar{g}:\R^n\rightarrow\R$ in the dynamics~\eqref{eq:RMPC_assumptions}.
\end{theorem}

\subsection{\RAMPC{} optimal control problem}\label{subsec:ocp_RAMPC}

Given a GP model collection~$\M_{k}$ and a selection of GPs~$\I_k$ at time~$t$, the OCP of the \RAMPC{} scheme is defined as follows:
%
\begin{subequations}\label{eq:RAMPC_optimization_problem}
    \begin{align}
        \min_{\substack{v_{\cdot|t},z_{\cdot|t},\\\delta_{\cdot|t},\{\lambda_{\cdot|t,i}\}_{i\in \I_k}}}& \; \int_{0}^{T_\mathrm{f}}\ell(z\traj,v\traj,\bg\traj) \dtau
        + \ell_\mathrm{f}(z_{T_\mathrm{f}|t},\bg_{T_\mathrm{f}|t})
        \\
        \text{s.t. }\quad 
        & \eqref{subeq:RMPC_formulation_constraint_1}-\eqref{subeq:RMPC_formulation_constraint_2}, \eqref{subeq:RMPC_formulation_constraint_3}-\eqref{subeq:RMPC_formulation_constraint_6}
        \\
        \label{subeq:RAMPC_formulation_constraint_2_bis}
        &\bg\traj(\cdot) = \tilde{g}(\cdot,\I_k,\lambda\traj),
        \\
        \label{subeq:RAMPC_formulation_constraint_2_w}
        & w\traj(\cdot) = \tilde{w}(\cdot, \bg\traj, \I_k).
    \end{align}
\end{subequations}

The main difference compared to~\eqref{eq:RMPC_optimization_problem} is the use of the estimate~$\bg$~\eqref{eq:bar_g_as_weighted_sum} and the uncertainty bound~$w$~\eqref{eq:def_w_tilde_RAMPC} in the respective constraints~\eqref{subeq:RAMPC_formulation_constraint_2_bis} and \eqref{subeq:RAMPC_formulation_constraint_2_w}, which both depend on multiple GP models and the additional decision variables~$\{\lambda_{\cdot|t,i}\}_{i\in\I_k}$. 
Analogous to the nominal input $v$, the decision variables $\lambda$ are piece-wise constant.

Algorithms~\ref{alg:offline_design_RMPC} and~\ref{alg:online_design_RAMPC} describe the offline and online computations of the \RAMPC{} scheme. In contrast to the \RMPC{} scheme, Algorithm~\ref{alg:online_design_RAMPC} includes multiple GPs, conditioned on data gathered online. Algorithm~\ref{alg:GP_selection} shows how the selection of GP models, expressed by the set~$\I_k$, is performed. 
This selection mechanism reduces the number of GPs and thus the computational demand that arises in practice when storing and evaluating multiple GPs, while preserving the closed-loop guarantees. 
More recent GPs contain more data points and are thus typically helpful in improving performance. 

\begin{algorithm}[t]\caption{Online computations (\RAMPC{}).}\label{alg:online_design_RAMPC}
\begin{algorithmic}[1]
    \For{\textnormal{each sampling time} $t_k=kT_s, k\in \mathbb{I}_{\geq 0}$}
    \State
    Solve the OCP~\eqref{eq:RAMPC_optimization_problem}.
    \State
    Apply the feedback~\eqref{eq:RMPC_control_input_from_ocp} over $t\in [t_k,t_{k+1})$.
    \State  \parbox[t]{\dimexpr\linewidth-\algorithmicindent}{%
    Obtain noisy measurement~\eqref{eq:noisy_measurement} at $t=t_{k+1}$ and store it in $\mathcal{D}_{t_{k+1}}$ (Sec.~\ref{subsec:online_measurements_RAMPC}).
    }
    \State
    Create updated GP~$\GP_{{k+1}} = \GP(\mathcal{D}_{t_{k+1}})$.
    \State  \parbox[t]{\dimexpr\linewidth-\algorithmicindent}{%
    Update GP model selection $\I_{k+1}$ using Alg.~\ref{alg:GP_selection}.
    }
    \EndFor
\end{algorithmic}
\end{algorithm}

    
    
    

        
        
        
            
            

    

\begin{algorithm}[t]\caption{GP model selection.}\label{alg:GP_selection}
\begin{algorithmic}[1]
    \For{$i\in\mathbb{I}_{k}$}
    
    \If{$\lambda^\star_{\tau,i}=0 \;\forall \tau\in [0,T_f]$ \textbf{and}~$i$ is not active when\\~~\quad evaluating the uncertainty bound~$\tilde{w}$~\eqref{eq:def_w_tilde_RAMPC} for the\\~~\quad optimal solution of~\eqref{eq:RAMPC_optimization_problem}}
        \State
        \parbox[t]{\dimexpr\linewidth-\algorithmicindent}{%
        Eliminate GP model $i$ from selection~$\I_k$,\\ i.e., $\I_{k} \leftarrow \I_k\setminus \{i\}$.
        }
    \EndIf
        
    \EndFor
    \State Add $\GP_{k+1}$ to selection, i.e., $\I_{k+1}=\I_{k} \cup \{k+1\}$.
\end{algorithmic}
\end{algorithm}

\subsection{Theoretical properties \& discussion}\label{subsec:RAMPC_theoretical_properties}
The following theorem shows that the \RAMPC{} scheme provides the same theoretical properties as the \RMPC{} scheme.
\begin{theorem}\label{thm:RAMPC_feasibility_convergence}
    Let Assumptions~\ref{assumption:measurement_noise_sequence_g_RKHS},~\ref{assumption:CCM}, and~\ref{assumption:terminal_set} hold. Suppose that the optimization problem~\eqref{eq:RAMPC_optimization_problem} is feasible at~$t_0=0$ for the initial state~$x_0$, the initial GP model collection~$\M_{0}=\{\GP(\mathcal{D}_{t_0})\}$, and the GP model selection~$\I_0 = \{0\}$. 
 Then, with at least probability $1-p$, the closed-loop system~\eqref{eq:system_definition} with input~\eqref{eq:RMPC_control_input_from_ocp} resulting from Algorithm~\ref{alg:online_design_RAMPC} 
    ensures the same  closed-loop properties as Theorem~\ref{thm:RMPC_feasibility_convergence}: recursive feasibility, constraint satisfaction~\eqref{eq:state_input_constraints}, and convergence~\eqref{eq:convergence_RMPC}, i.e., for~$~\tau\in[0,T_\mathrm{s})$,
    \begin{equation}
       \lim_{k\rightarrow \infty}\norm{(z^\star_{\tau|t_k},v^\star_{\tau|t_k})-(x_\mathrm{ref}(\bg^\star\traj),u_\mathrm{ref}(\bg^\star\traj))}=0.
    \end{equation}
\end{theorem}
The proof of Theorem~\ref{thm:RAMPC_feasibility_convergence} is analogous to Theorem~\ref{thm:RMPC_feasibility_convergence}, using the fact that we can keep a consistent prediction model~\eqref{eq:g_bar_k_plus_one_g_bar} and the uncertainty bounds are monotone~\eqref{eq:set_inclusion_w}. 
Existing GP-MPC approaches cannot ensure recursive feasibility with online updates of the GP model ~\cite{koller2018learning,prajapat2025sampling}, or only if the model update is subject to an additional feasibility check~\cite{maiworm2021online,cao2025optimistic}. 
In contrast, the proposed approach seamlessly integrates online data with an updated GP model $\GP_k$ by using set intersection for uncertainty bounds (Prop.~\ref{prop:bound_g_g_bar}) and optimizing over the nominal dynamics using the linear combination~\eqref{eq:bar_g_as_weighted_sum}. 
This approach is related to the handling of finite-dimensional parametric uncertainties in~\cite{sasfi2023RAMPC}, which also involves set intersections and freely optimized nominal parameters. 
Importantly, the proposed approach integrates these features through finite-dimensional optimization variables $\lambda$, despite the (infinite-dimensional) non-parametric nature of GPs.

The high-probability GP estimation bound in Lemma~\ref{lemma:GP_estimate_uncertainty_bound} 
requires that the GP models $\mathcal{GP}_k$ use nested data sets $\mathcal{D}_{t_k} \subseteq \mathcal{D}_{t_{k+1}}$. 
    As the computational complexity of GP inference scales with the size of the data set~\cite{scampicchio2025gaussian}, 
    there is a limit on the
    maximum number of new data points
    that can 
    be added online
    for a given computational budget.
    This issue can be addressed by only using GPs with a maximal number of data points (cf. Sec.~\ref{sec:implementation_details}), which does not affect the guarantees in Theorem~\ref{thm:RAMPC_feasibility_convergence}.

\section{Numerical example} \label{sec:numerical_example}

In this section, we compare the \RAMPC{} scheme to the \RMPC{} scheme by applying them to a planar quadrotor model adapted from~\cite{zhao2022tube,sasfi2023RAMPC}\ifbool{arxiv}{, which is depicted in Figure~\ref{fig:quadrotor}}
{%
}. \ifbool{arxiv}{Section~\ref{sec:quadrotor_setup} describes the quadrotor's dynamics and the control task, Section~\ref{sec:implementation_details} presents implementation details, and Section~\ref{sec:numerical_results} shows the obtained numerical results.}
{
}
The code is available online\footnote{Source code: \texttt{https://gitlab.ethz.ch/ics/gp-rampc}, \texttt{10.3929/ethz-c-000803178}.}; additional details on the setup and implementation can be found in~\cite{dubied2025robust}.

\subsection{Setup}\label{sec:quadrotor_setup}
\begin{figure}[t]
    \centering
    \includegraphics[trim=0 5 0 5,clip,width=\linewidth]{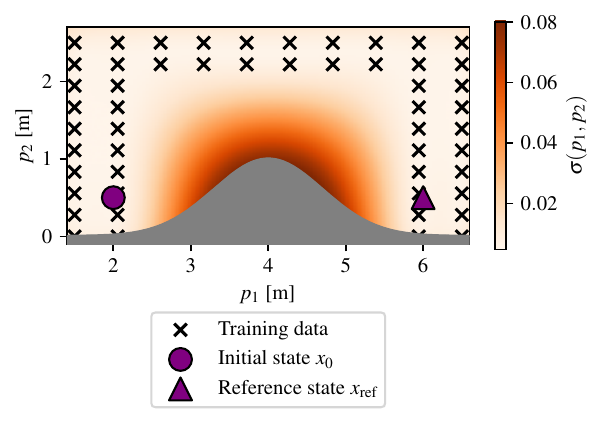}
    \caption{Environment, including  offline training data. Standard deviation of the GP is shown with a colormap.}
    \label{fig:quadrotor_setup}
\end{figure}

\ifbool{arxiv}{
\begin{figure}[t]
    \centering
    \includegraphics[width=0.9\linewidth]{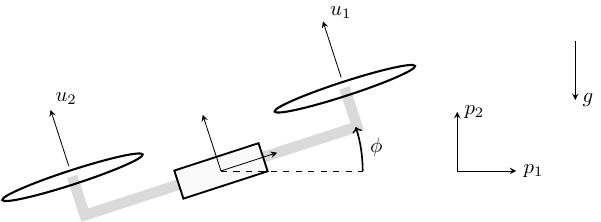}
    \caption{Planar quadrotor.}
    \label{fig:quadrotor}
\end{figure}
}{
}
The state of the system is~$x=[p_1,p_2,\phi,v_1,v_2,\dot{\phi}]^\top\in\mathbb{R}^6$, with~$p_1$,~$p_2$ describing its horizontal and vertical positions,~$\phi$,~$\dot{\phi}$ its angle and angular velocity, and~$v_1$,~$v_2$ the velocities in the body frame of the quadrotor. The \ifbool{arxiv}{}{exact} dynamics of the system are described
\ifbool{arxiv}{ by
{\allowdisplaybreaks\begin{align}
        &f_w(x,u,g,d) 
        \nonumber
        \\
        =& 
        f(x) + B(x)u + Gg(x) + E(x)d
        \nonumber
        \\
        =&
        \begin{bmatrix}
             v_1\cos(\phi) - v_2\sin(\phi)  \\
             v_1\sin(\phi) + v_2\cos(\phi)   \\
             \dot{\phi}                     \\
             v_2\dot{\phi} - g\sin(\phi)    \\
             -v_1\dot{\phi} - g\cos(\phi)   \\
             0
        \end{bmatrix}
        +
        \begin{bmatrix}
            0   &   0   \\
            0   &   0   \\
            0   &   0   \\
            0   &   0   \\
            \frac{1}{m} &   \frac{1}{m} \\
            \frac{l}{J} &   \frac{-l}{J}  
        \end{bmatrix}
        \begin{bmatrix}
            u_1     \\
            u_2
        \end{bmatrix}
        \nonumber
        \\
        &+
        \begin{bmatrix}
           0    \\
           0    \\
           0    \\
           0    \\
           1    \\
           0      
        \end{bmatrix}
            \underbrace{\frac{0.25}{3.0\cdot  \mathrm{height}(p_1,p_2)+1}}_{g(x)}
        +
        \begin{bmatrix}
            0   \\
            0   \\
            0   \\
            \cos(\phi)  \\
            -\sin(\phi) \\
            0
        \end{bmatrix}
        d,
        \nonumber
\end{align}}%
}{in \cite{dubied2025robust}.}
\ifbool{arxiv}{
where~$m=0.486$ \si{kg} is the mass of the system,~$J=0.00383$ \si{kg/m^2}, its moment of inertia,~$l=0.25$ \si{m}, the distance between the center of the quadrotor and its propellers, and~$d$, a random wind disturbance. The control input~$u=[u_1,u_2]^\top$ is the force stemming from the propellers. Compared to the setup presented in~\cite{zhao2022tube,sasfi2023RAMPC}, the quadrotor is also subject to an unknown vertical force~$g(x)$ due to ground effects depending on the vertical distance to the ground~$\mathrm{height}(p_1,p_2)$. The magnitude of this force is decreasing with increasing distance to the ground. Figure~\ref{fig:quadrotor_setup} shows the environment characterized by the presence of a hill, the available offline measurements, and the resulting standard deviation of the initial GP. The GP has a low variance around the initial state~$x_0$ and reference state~$\xref$, as well as in the region further away from the ground. In contrast, the uncertainty is larger in the region around the hill.}
{
 Compared to the setup presented in~\cite{zhao2022tube,sasfi2023RAMPC}, the quadrotor is also subject to an unknown vertical force~$g(x)$ due to ground effects depending on the vertical distance to the ground~$\mathrm{height}(p_1,p_2)$:
 \begin{equation*}
     g(x)=\frac{0.25}{3.0\cdot  \mathrm{height}(p_1,p_2)+1}, \quad G=[0,0,0,0,1,0]^\top.
 \end{equation*} 
 The magnitude of this force is decreasing with increasing distance to the ground. Figure~\ref{fig:quadrotor_setup} shows the environment characterized by the presence of a hill, the available offline measurements, and the resulting standard deviation of the initial GP. The GP has a low variance around the initial state~$x_0$ and reference state~$\xref$, as well as in the region further away from the ground. In contrast, the uncertainty is larger in the region around the hill.
}
\ifbool{arxiv}{
The following constraints are used:
\begin{subequations}
    \begin{align*}
        (v_1,v_2) &\in [-2,2]\times [-1,1] \;\text{\si{m/s}},\\
        (\phi,\dot{\phi}) &\in [-22.5\degree,22.5\degree]\times [-60,60] \;\text{\si{\degree/s}},\\
        d&\in \D = [-0.02,0.02],\\
        (u_1,u_2)&\in [-1,3.5]\times[-1,3.5],
    \end{align*}
\end{subequations}
in addition to a nonlinear constraint on~$p_1$ and~$p_2$ enforcing collision avoidance with the ground.
\\
}
{
}
The control task is to fly from the initial state~$x_0$ to the reference state~$\xref$, as shown in Figure~\ref{fig:quadrotor_setup}. \ifbool{arxiv}{The stage cost~\eqref{eq:def_quadratic_stage_cost} is defined by 
    $Q_c=\text{diag}(20,20,10,1,1,10)$ and
    $R_c=\text{diag}(5,5)$.
}{}

\begin{figure}[t]
    \centering
    \includegraphics[trim=0 7 0 5,clip,width=\linewidth]{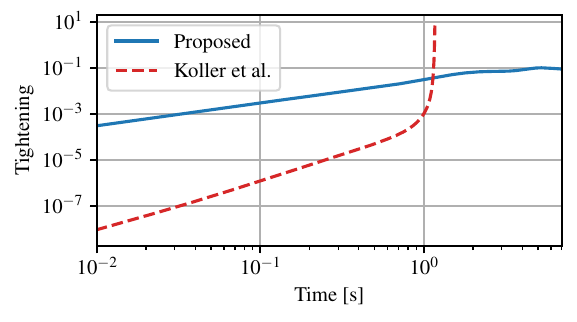}
    \caption{Comparison between the size of the predicted reachable set obtained using the proposed contraction-metric-based tube construction and existing work~\cite{koller2018learning}. The constraint tightening is computed as~$c_1\delta_{t|0}$\eqref{subeq:RMPC_formulation_constraint_3}, corresponding to the constraint for the state~$p_1$.}
    \label{fig:Koller_comparison}
\end{figure}

\subsection{Implementation details}\label{sec:implementation_details}
\ifbool{arxiv}{
All computations are performed on a Lenovo Thinkpad T14, equipped with an AMD Ryzen 7 PRO 6850U processor and 32GB of RAM, using Windows Subsystem for Linux.
First, the contraction metrics (Asm.~\ref{assumption:CCM}) are computed using LMIs in Algorithm~\ref{alg:offline_design_RMPC}, 
which are modified from the code in \cite{erdin2024RMPC} based on~\cite{zhao2022tube,sasfi2023RAMPC} 
and uses \texttt{YALMIP}~\cite{lofberg2004yalmip} 
and
\texttt{Mosek}~\cite{andersen2000mosek}.
The set~$\mathbb{S}$ (Asm.~\ref{assumption:measurement_noise_sequence_g_RKHS}) is computed as a hyperbox by considering the maximum of the partial derivatives of the true function~$g(x)$ for~$x\in\Z_x$. 
The constant $\beta_N$~\eqref{eq:uncertainty_bound_fiedler} depends on the bound~$B_g$ on the RKHS norm (Asm.~\ref{assumption:measurement_noise_sequence_g_RKHS}); similar to other GP-MPC implementations (cf.~\cite{koller2018learning,prajapat2025safe}), we assume $\beta_N=3.0$. 
For the considered dynamics, the reference $\xref$ is independent of the estimate $\bar{g}$. Thus, the decision variable $\bar{\delta}_\mathrm{f}$ in the terminal set design (Prop.~\ref{prop:terminal_set}) is computed offline, by considering the range of $\uref(\bar{g})$ for the given bound $|\bar{g}(x)|\leq 0.25$. The offline computations took 5.5~\si{min}.

We discretize the dynamics using a fourth-order Runge-Kutta integrator with a step size of~$T_\mathrm{s}=0.15$~\si{s} and set the horizon length to~$N=50$,  ($T_\mathrm{f}=7.5$~\si{s}).
The OCPs~\eqref{eq:RMPC_optimization_problem} and \eqref{eq:RAMPC_optimization_problem} are solved using \texttt{L4acados}~\cite{lahr_l4acados_2024}, which employs \texttt{acados}~\cite{Verschueren2021acados} and \texttt{GPyTorch}~\cite{gardner2018gpytorch}. 
The maximum number of sequential quadratic programming (SQP) iterations in \texttt{acados} is set to 50. For the closed-loop simulation, we apply Alg.~\ref{alg:online_design_RAMPC} for 35 time steps steps and use soft constraints to ensure feasibility of intermediate QP solves. 
  
During the online operation, the RAMPC scheme (Alg.~\ref{alg:online_design_RAMPC}) needs to evaluate several GP models stored in the GP model collection~$\M_{k}$~\eqref{eq:def_M_tk}. 
Fixed runtime is achieved by evaluating GPs in a batch consisting of a fixed maximal number of~$n_b=10$ GPs and~$n_\mathrm{mobs}=200$ data points, where yet unused input locations are masked out in the \texttt{GPyTorch} implementation. 
At initialization, the GPs use $N=52$ offline data point (see Fig.~\ref{fig:quadrotor_setup}).
The proposed implementation achieves a fixed runtime but yields two limitations. 
As the number of stored GPs in the batch is capped to $n_b$, we only use an updated GP if the GP model selection (Alg.~\ref{alg:GP_selection}) is effective. 
Furthermore, once the maximal number of $n_{\mathrm{mobs}}$ data points is reached, the GP model is no longer updated. 
}{The contraction metrics (Asm.~\ref{assumption:CCM}) are computed using LMIs in Algorithm~\ref{alg:offline_design_RMPC}, which are modified from the code in \cite{erdin2024RMPC} based on~\cite{zhao2022tube,sasfi2023RAMPC} and use \texttt{YALMIP}~\cite{lofberg2004yalmip}  and \texttt{Mosek}~\cite{andersen2000mosek}.
We discretize the dynamics using a fourth-order Runge-Kutta integrator with a step size of~$T_\mathrm{s}=0.15$~\si{s} and set the horizon length to~$N=50$,  ($T_\mathrm{f}=7.5$~\si{s}).
The OCPs~\eqref{eq:RMPC_optimization_problem} and \eqref{eq:RAMPC_optimization_problem} are solved using sequential quadratic programming (SQP) in \texttt{L4acados}~\cite{lahr_l4acados_2024}, which solves a series of quadratic problems (QP) using \texttt{acados}~\cite{Verschueren2021acados} and \texttt{GPyTorch}~\cite{gardner2018gpytorch}. 
}

\subsection{Results}\label{sec:numerical_results}
In Figure~\ref{fig:Koller_comparison}, we compare the size of the predicted tube of the proposed \RMPC{} formulation and the sequential ellipsoidal propagation proposed in~\cite{koller2018learning}. 
The method in~\cite{koller2018learning} uses a Taylor expansion and the accumulating linearization error yields an exponential increase of the tube, resulting in numerical divergence after approximately 1~\si{s}. 
In contrast, the proposed approach only requires to simulate the scaling $\delta$ using the offline computed contraction metric and the error remains bounded.

\begin{figure}[t]
    \centering
    \includegraphics[trim=0 5 0 5,clip, width=\linewidth]{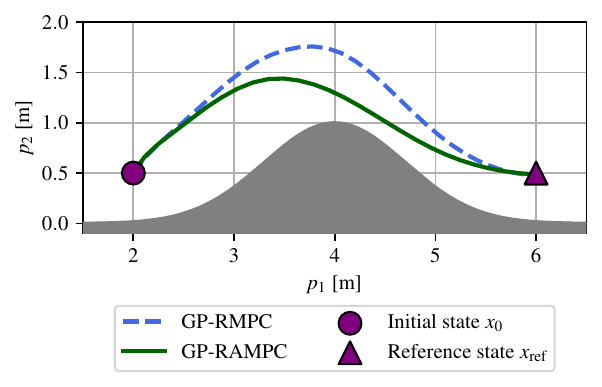}
    \caption{Comparison between the quadrotor's trajectories obtained by applying the \RMPC{} and the \RAMPC{} schemes.}
    \label{fig:RAMPC_RMPC_comparison}
\end{figure}

Figure~\ref{fig:RAMPC_RMPC_comparison} compares the \RMPC{} and \RAMPC{} schemes. 
The \RAMPC{} scheme uses data obtained during runtime to reduce uncertainty, allowing the controller to reach the terminal set $6\%$ faster and reducing the closed-loop tracking cost by $9\%$ compared to the \RMPC{}. The following table lists the average computation times per SQP iteration, which are dominated by the GP evaluation and approximately constant during runtime due to the proposed implementation strategy.
\begin{center}
\begin{tabular}{l|c|c|c}
Scheme & GP Comp. & QP Solve & Total \\\hline
\RMPC{} & 45.6~\si{ms}  & 5.5~\si{ms} & 51.2~\si{ms} \\
\RAMPC{} & 79.5~\si{ms} & 9.5~\si{ms}& 89.1~\si{ms}
\end{tabular}
\end{center}

%

\begin{figure}[t]
    \centering
    \includegraphics[trim=0 5 0 0,clip,width=\linewidth]{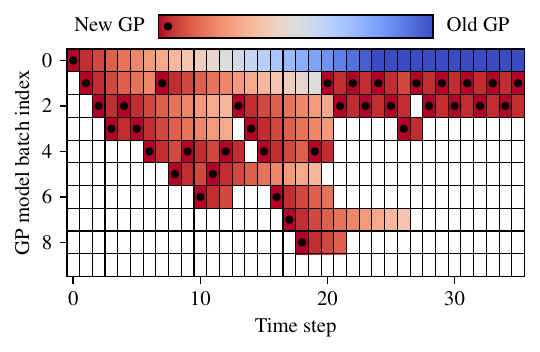}
    \caption{Evolution of the GP models stored across the different batch dimensions. Addition of a new GP model to the selection~$\I_k$ according to Alg.~\ref{alg:GP_selection} is represented by a black dot. If a GP model cannot be removed according to Alg.~\ref{alg:GP_selection}, it stays in the selection~$\I_k$. The quadrotor enters the terminal set at time step 31.}
    \label{fig:batch_map}
\end{figure}

Finally, Figure~\ref{fig:batch_map} shows the time evolution of the GP models saved in the batch according to the selection $\I_k$ in Algorithm~\ref{alg:GP_selection}. 
Note that the maximum number of models  $n_b=10$ is never reached. Additionally, the new GP model at each step $k$ is utilized, taking all online measurements into consideration (as represented by the red cells with the black dot).

\section{Conclusion} \label{sec:conclusion}
We have presented a GP-MPC formulation, which, with a user-specified probability,  ensures recursive feasibility and constraint satisfaction for all times by using contraction metrics and high-probability bounds for GPs to provide efficient robust predictions. 
The presented robust adaptive GP-MPC formulation uses online data to adapt the model, which improves performance while recursive feasibility is ensured by enabling the optimizer to freely interpolate between a collection of GP models. 
A numerical example demonstrates the benefits of the proposed robust prediction and online adaptation.
Open issues include enhancing numerical efficiency by using tailored optimization methods~\cite{lahr2023zero} and investigating more advanced data management strategies~\cite[Sec.~4.1]{scampicchio2025gaussian}.

\bibliographystyle{plain}        
\bibliography{Literature2}

\ifbool{arxiv}{\appendix

\allowdisplaybreaks}

\ifbool{arxiv}{
\section{Details of the Proofs} \label{sec:proofs}
First,
we present the proofs of the theoretical results of Section~\ref{sec:RAMPC_framework}. 
We start by presenting a simple constructive design of the terminal set that satisfies Assumption~\ref{assumption:terminal_set} (Prop.~\ref{prop:terminal_set}) in Appendix~\ref{sec:proof_terminal_set}.
Then, we provide preparatory results used in the proof of Theorem~\ref{thm:RAMPC_tube}:
\begin{itemize}
    \item Appendix~\ref{sec:preparatory_results}: Results for contraction metrics  (Prop.~\ref{prop:Lyapunov_function_dynamics}, Lemmas~\ref{lemma:offline_constant_L_G} and~\ref{lemma:constraint_satisfaction});
    \item Appendix~\ref{sec:proof_uncertainty_bound}: Uncertainty bound for GP model collection~$\M$ (Prop.~\ref{prop:bound_g_g_bar}).
\end{itemize}

Then, we provide the following main proofs:

\begin{itemize}
    \item Appendix~\ref{sec:proof_constraint_satisfaction_trajectory}: Robust reachability using contraction metrics
     (Thm.~\ref{thm:RAMPC_tube});
    \item Appendix~\ref{sec:proof_theoretical_properties}: Closed-loop properties of the RAMPC scheme (Thm.~\ref{thm:RAMPC_feasibility_convergence});
     \item Appendix~\ref{sec:proof_RMPC_special_case}: Discussion of the \RMPC{} scheme as a special case of the \RAMPC{} case (Thm.~\ref{thm:RMPC_tube} and~\ref{thm:RMPC_feasibility_convergence}).
\end{itemize}

All theoretical results leverage the uncertainty bound from Lemma~\ref{lemma:GP_estimate_uncertainty_bound}, which holds jointly for all times and input locations with a probability of at least $1-p$; hence, all the intermediate claims are valid with the same probability.
}{}

\ifbool{arxiv}{
\subsection{Terminal set constructive design}\label{sec:proof_terminal_set}
In this section, we provide a simple constructive design satisfying Assumption~\ref{assumption:terminal_set}.
\begin{proposition}\label{prop:terminal_set}
    Let Assumption~\ref{assumption:CCM} hold. Then, Assumption~\ref{assumption:terminal_set} is satisfied with zero terminal cost ($\ell_\mathrm{f}=0$), reference input~$v_\mathrm{f}=\uref(\bg)$, and the following terminal set
    \begin{subequations}\label{eq:terminal_set_if_remark}
        \begin{align}
            \Xterminal = \bigg\{
            &(z,\delta,\bg,w)\bigg| 
            \nonumber
            \\
            \label{eq:terminal_set_b}
            & \exists\bar{\delta}_\mathrm{f}\in \R_{\geq 0}\;:\; \delta\in [0,\bar{\delta}_\mathrm{f}],
            \\
            \label{eq:terminal_set_a}
            &
            z=\xref(\bg), 
            \\
            \label{eq:terminal_set_c}
            & f_\delta(z,\bar{\delta}_\mathrm{f},w) \leq 0,
            \\
            \label{eq:terminal_set_d}
            & h_j(z,\uref(\bg)) + c_j \bar{\delta}_\mathrm{f} \leq 0,\; \forall j\in \Nr
           \bigg\}.
        \end{align}
    \end{subequations}
\end{proposition}
\begin{pf}
The result is an adaptation of ~\cite[Prop.~11]{sasfi2023RAMPC}. 
Applying $v_\mathrm{f}=\uref(\bg)$ in combination to condition~\eqref{eq:terminal_set_a} yields $z_\tau=\xref(\bg)$ for~$\tau\in[0,T_\mathrm{s}]$. In addition, conditions~\eqref{eq:terminal_set_b} and~\eqref{eq:terminal_set_c} ensure that $\delta_\tau\in [0,\bar{\delta}_\mathrm{f}]$ for~$\tau\in[0,T_\mathrm{s}]$ and therefore the positive-invariance condition~\eqref{eq:terminal_set_PI} holds. Furthermore, the fact that~$\delta\in [0,\bar{\delta}_\mathrm{f}]$ for~$\tau\in[0,T_\mathrm{s}]$, in combination with~\eqref{eq:terminal_set_d}, yields constraint satisfaction~\eqref{eq:terminal_set_constraint_satisfaction}. The decrease conditions~\eqref{eq:terminal_set_local_CLF} follows from~$\ell_\mathrm{f}=0$ and the stage cost~\eqref{eq:def_quadratic_stage_cost}. The monotonicity property~\eqref{eq:terminal_set_monotonicity} follows from Condition~\eqref{eq:terminal_set_b} and the monotonicity of  of~$f_\delta$~\eqref{subeq:RMPC_assumption_4_part2} in Condition~\eqref{eq:terminal_set_c}, yielding~$f_\delta(z,\bar{\delta}_\mathrm{f},w-\hat{w})\leq f_\delta(z,\bar{\delta}_\mathrm{f},w)$.
 $~\hfill \square$
\end{pf}
}{}

\ifbool{arxiv}{
\subsection{Results for contraction metrics}\label{sec:preparatory_results}
In this section, we derive bounds for contraction metrics with GP models, which are used to prove Theorem~\ref{thm:RAMPC_tube}. 
The following lemma derives a Lipschitz-continuity-like bound that is later used in Proposition~\ref{prop:Lyapunov_function_dynamics}. 
\begin{lemma}\label{lemma:offline_constant_L_G}
     Suppose Assumption~\ref{assumption:measurement_noise_sequence_g_RKHS} holds. Then, for any~$x,z\in \Z_x$ such that~$\gOpt\in\Z_x$ for~$s\in[0,1]$, the following bound holds: \begin{equation}\label{eq:bound_norm_Gg_LG}
        \norm{\Mgammahalf{s} G(g(x)-g(z))}
        \leq 
        L_G \Vdelta(x,z),
    \end{equation}
with
\begin{equation}
            L_G
            :=\max_{x,x'\in \Z_x}\max_{g_x\in\mathbb{S}}\left\{\left\Vert M(x)^{\frac{1}{2}}Gg_x M(x')^{-\frac{1}{2}}\right\Vert\right\}. 
\end{equation}
   
\end{lemma}
}{}
\ifbool{arxiv}{
\begin{pf}
    It holds that
    \begin{align*}
    &\norm{\Mgammahalf{s} G(g(x)-g(z))}
            \\
            =
            &
            \left\Vert\Mgammahalf{s} G\int_0^1 \pder{g}{x}\bigg|_{\gamma^\star(s')}\gamma_s^\star(s')\ds'\right\Vert
            \\
            =
            &
            \norm{\Mgammahalf{s} G
            \nonumber
            \\
            &
            \int_0^1 \pder{g}{x}\bigg|_{\gamma^\star(s')}\Mgammamhalf{s'}\Mgammahalf{s'}\gamma_s^\star(s')\ds'}\\
            \leq &\int_0^1\left\|\Mgammahalf{s} G\pder{g}{x}\bigg|_{\gamma^\star(s')}\Mgammamhalf{s'}\right\|\\ &\cdot \norm{\Mgammahalf{s'}\gamma_s^\star(s')}\ds'\\
            \leq &\max_{s'\in[0,1]}\left\|\Mgammahalf{s} G\pder{g}{x}\bigg|_{\gamma^\star(s')}\Mgammamhalf{s'}\right\|\\         & \cdot   \int_0^1\norm{\Mgammahalf{s'}\gamma_s^\star(s')}\ds'\\
            \stackrel{\mathmakebox[\widthof{=}]{\eqref{eq:def_Lyapunov_function}}}{\leq }&\max_{x,x'\in \Z_x}\max_{g_x\in\mathbb{S}}\left\|M(x)^{1/2} G g_x M(x')^{-1/2}\right\| V_\delta(x,z)\\
            \stackrel{\eqref{eq:def_L_G}}{=} & L_G V_\delta(x,z),
            \end{align*}
where the first equation uses the gradient theorem and the first inequality uses the triangular inequality for integrals and the induced matrix norm.
$\hfill \square$
\end{pf}
}{}
%
%
\ifbool{arxiv}{
The following proposition establishes a robust bound on the derivative of the incremental Lyapunov function.
\begin{proposition}\label{prop:Lyapunov_function_dynamics}
    Suppose Assumptions~\ref{assumption:measurement_noise_sequence_g_RKHS} and~\ref{assumption:CCM} hold. Consider a GP model selection~$\I_{k}$, an arbitrary function~$\bg:\R^n\rightarrow \R$, and the corresponding bound~$w(\cdot)=\tilde{w}(\cdot,\bg, \I_k)$ according to~\eqref{eq:def_w_tilde_RAMPC}.
    Then, for all~$x,z\in \Z_x$ such that~$(\gOpt,\gammaU)\in\Z$, 
     $\forall s\in[0,1]$,  the following bound holds with a probability of at least~$1-p$:
    \begin{align*}
        &\dot{\Vdelta} (x,z)
        =
        \pder{\Vdelta}{x}\bigg|_{(x,z)} \xd + \pder{\Vdelta}{z}\bigg|_{(x,z)} \zd
        \\
        &
        \leq
        -\left(\rho - L_G \right)\Vdelta(x,z)
        + G_M w(z)+ E_M,
    \end{align*}
    with~$\dot{x}=f_w(x,u,g,d)$,~$\dot{z}=\bar{f}(z,v,\bg)$, $w(\cdot)=\tilde{w}(\cdot,\bg,\I_k)$, $u=\kappa(x,z,v)$, $L_G\geq 0$ according to~\eqref{eq:def_L_G}, and where
    \begin{subequations} 
        \begin{align}
            G_M
            & :=
            \max_{x\in \Z_x}\left\{ \norm{G}_{M(x)} \right\},
            \\
            E_M 
            & := 
            \max_{x\in \Z_x,\, d\in \D}\left\{\norm{E(x)d}_{M(x)}\right\}.
        \end{align}
    \end{subequations}
\end{proposition}
}{}

\ifbool{arxiv}{
\begin{pf}
    This proof is inspired by~\cite[Prop.~2]{sasfi2023RAMPC}, which is itself based on theory of contraction metrics~\cite{manchester2017control,zhao2022tube}. First, we define the dynamics of the geodesic:
    \begin{subequations}\label{eq:def_all_gamma}
        \begin{align}
            \gOptd 
            &
            = f_w(\gOpt,\gammaU,
            \bg(z)+s(g(x)-\bg(z)),sd),
            \nonumber
            \\
            \label{subeq:def_dot_gamma_opt_s}
            \gOptsd
            &
            = \Aclo(\gOpt,\gammaU,
            \bg(z)
            \nonumber
            \\
            & \phantom{=}
            +s(g(x)-\bg(z)),sd)\gOpts + \gammaW,
            \\
            \label{subeq:def_gamma_w}
            \gammaW
            &
            := G\left(g(x)-\mu_N(z)\right) + E(\gOpt)d.
        \end{align}
    \end{subequations}
    Note that $\dot{\gamma}^\star(0)=\dot{z}$, $\dot{\gamma}^\star(1)=\dot{x}$, and \eqref{subeq:def_dot_gamma_opt_s} is simply its derivative with respect to~$s$.
    For the sake of convenience, we denote by~$\Aclo(s)$ the first term on the right-hand-side of~\eqref{subeq:def_dot_gamma_opt_s}. In addition, the Riemannian energy is defined as 
    \begin{equation}\label{eq:def_Riemannian_energy}
        \mathcal{E}(x,z):=\int_0^1 \gOpts^\top \Mgamma{s}\gOpts \ds,
    \end{equation}
    and satisfies~$\mathcal{E}(x,z) = \Vdelta(x,z)^2$~\cite{manchester2017control,zhao2022tube}.  
   The dynamics of the Riemannian energy satisfy
        {\allowdisplaybreaks\begin{align*}
            \dot{\mathcal{E}}(x,z)
            =
            &
            \int_0^1 
            \gOpts^\top \dot{M}(\gamma^\star(s)) \gOpts
            \nonumber
            \\
            &
            +
            \gOptsd^\top \Mgamma{s}\gOpts 
            \nonumber
            \\
            &
            + \gOpts^\top \Mgamma{s}\gOptsd \ds 
            \\
            \stackrel{\mathmakebox[\widthof{=}]{\eqref{subeq:def_dot_gamma_opt_s}}}{=}
            \;\;
            &
            \int_0^1 
            \gOpts^\top \dot{M}(\gamma^\star(s)) \gOpts
            \nonumber
            \\
            &
            + \left(\Aclo(s) \gOpts +\gammaW\right)^\top \Mgamma{s} \gOpts 
            \nonumber
            \\
            &
            + \gOpts^\top \Mgamma{s}\left(\Aclo(s) \gOpts \right.
            \nonumber
            \\
            &
            \left. + \gammaW\right)\ds
            \\
            \stackrel{\mathmakebox[\widthof{=}]{\eqref{subeq:M_Acl_bound}}}{\leq}
            &
            \int_0^1 -2\rho \gOpts^\top \Mgamma{s}\gOpts 
            \nonumber
            \\
            &
            + 2 \gOpts^\top \Mgamma{s} \gammaW \ds.
            \\
            \stackrel{\mathmakebox[\widthof{=}]{\eqref{eq:def_Riemannian_energy}}}{=}
            &
            \;-2\rho \mathcal{E}(x,z) 
            \nonumber
            \\
            &
            + \int_0^1 2 \gOpts^\top \Mgamma{s} \gammaW \ds.
        \end{align*}}
    We further expand the second term to obtain 
    {\allowdisplaybreaks\begin{align}
        &\int_0^1 2\gOpts^\top \Mgamma{s} \gammaW \ds
        \nonumber
        \\
        =& \int_0^1 2 \gOpts^\top \Mgammahalf{s} \Mgammahalf{s} \gammaW \ds
        \nonumber
        \\
        \leq & 
        \int_0^1 2\norm{\Mgammahalf{s}\gOpts} \norm{\Mgammahalf{s} \gammaW} \ds
        \nonumber
        \\
        \leq &
        2\max_{s\in[0,1]} \norm{\Mgammahalf{s} \gammaW}\int_0^1 \norm{\Mgammahalf{s}\gOpts}  \ds
        \nonumber
        \\\label{subeq:expansion_gamma_s_M_gamma_w_part_4}
        \stackrel{\mathmakebox[\widthof{=}]{\eqref{eq:def_Lyapunov_function}}}{=} & \;
        2\Vdelta(x,z)\max_{s\in[0,1]} \norm{\Mgammahalf{s} \gammaW}.       
    \end{align}}%
    Thanks to the uncertainty bound $\tilde{w}$ from Proposition~\ref{prop:bound_g_g_bar} and the (Lipschitz) constant $L_G$ from Lemma~\ref{lemma:offline_constant_L_G}, we can bound the maximum term in~\eqref{subeq:expansion_gamma_s_M_gamma_w_part_4} as follows:
    {\allowdisplaybreaks\begin{subequations}
        \begin{align}
            &
            \max_{s\in[0,1]} \norm{\Mgammahalf{s} \gammaW}
            \;\;
            \nonumber
            \\
            \stackrel{\mathmakebox[\widthof{=}]{\eqref{subeq:def_gamma_w}}}{=}\;\;
            &
            \max_{s\in [0,1]} \norm{ \Mgammahalf{s}\left[G\left(g(x)-\bg(z)\right)  + E(\gOpt)d\right]}
            \nonumber
            \\
            \label{subeq:proof_tube_dynamics_last_part_3}
            \leq
            &
            \max_{s\in [0,1]}\left\{ \norm{ \Mgammahalf{s}G(g(x)-g(z))} \right.
            \nonumber
            \\
            & + \norm{\Mgammahalf{s}G(g(z)-\bg(z))}
            \nonumber
            \\
            & \left. + \norm{{\Mgammahalf{s}E(\gOpt)d}} \right\}
            \\
            \label{subeq:proof_tube_dynamics_last_part_4}
            \leq
            &
            \max_{s\in [0,1]}\left\{ \norm{ \Mgammahalf{s}G(g(x)-g(z))}\right\}
            \nonumber
            \\
            & + \max_{s\in [0,1]}\left\{\norm{\Mgammahalf{s}G}\right\}|g(z)-\bg(z)|
            \nonumber
            \\
            & + \max_{s\in [0,1]}\left\{\norm{{\Mgammahalf{s}E(\gOpt)d}} \right\}
            \nonumber
            \\
            \stackrel{\mathmakebox[\widthof{=}]{\substack{\eqref{eq:bound_norm_Gg_LG}\\\eqref{eq:prop_bound_g_g_bar}\eqref{eq:def_offline_constants}}}}{\leq}
            &
            \quad\;\; L_G \Vdelta(x,z) + G_M w(z) + E_M ,
        \end{align}
    \end{subequations}}%
    where~$L_G$,~$G_M$ and~$E_M$ are the constants defined in~\eqref{eq:def_offline_constants}, and~$w(\cdot)=\tilde{w}(\cdot,\bar{g},\I_k)$.     
    Bringing all the previous steps together, we obtain
    \begin{align*}
        \dot{\mathcal{E}}(x,z) \leq 
        &
        -2\rho \mathcal{E}(x,z)
        + 2 \Vdelta(x,z)
        \nonumber
        \\
        &
        \left( L_G\Vdelta(x,z)+ G_M w(z) + E_M \right).
    \end{align*}  
    Finally, because~$\mathcal{E}=\Vdelta^2$ and \mbox{$\dot{\mathcal{E}}=2\Vdelta\dot{\Vdelta}$}, we obtain
    \begin{equation*}
        \dot{\Vdelta} (x,z)
        \leq
        -\left(\rho - L_G \right)\Vdelta(x,z) 
        + G_M w(z) + E_M.~\hfill \square
    \end{equation*}   %
\end{pf}
}{}

\ifbool{arxiv}{
The following lemma describes sufficient conditions such that the geodesic lies in the constraint set.
\begin{lemma}\label{lemma:constraint_satisfaction} \cite[Prop.~5]{sasfi2023RAMPC}
    Suppose Assumption~\ref{assumption:CCM} holds. Then, for any~$x,z\in \R^n$,~$v\in\R^m$ satisfying
    \begin{subequations}
        \begin{align*}
            &
            h_j(z,v)+c_j\Vdelta(x,z)\leq0, \quad j\in\Nr,
            \\
            &
            c_j := \max_{(z,v)\in \Z }\left\Vert\left(\pder{h_j}{x}\bigg|_{(z,v)}+\pder{h_j}{u}\bigg|_{(z,v)} K(z)\right)M(z)^{-\frac{1}{2}}\right\Vert,
        \end{align*}
    \end{subequations}
    it holds that
    \begin{equation}\label{eq:geodesic_in_constraint_set}
        (\gOpt,\gammaU)\in \Z,\quad \forall s\in[0,1],
    \end{equation}
    where~$\gamma^\star$ and~$\gamma^u$ according to Eqs.~\eqref{eq:def_Lyapunov_function} and~\eqref{eq:def_feedback_input_part_2}.
\end{lemma}
}{}

\ifbool{arxiv}{
\subsection{Proof of Proposition~\ref{prop:bound_g_g_bar}}\label{sec:proof_uncertainty_bound}
Considering the uncertainty bound on the GP~\eqref{eq:prop_bound_g_g_bar}, we note that the GPs are build sequentially on the same data. This allows to use Lemma~\ref{lemma:GP_estimate_uncertainty_bound}, which holds jointly for all~$i\in\I_k$, i.e.,
\begin{align*}
    &\mathbb{P}[|g(z)-\mu_{\GP_i}(z)|\leq \beta_{\GP_i}\sigma_{\GP_i}(z),\; \forall z\in \Z_x,i\in \I_k]
    \\
    &\leq 1-p,
\end{align*}
which can be equivalently written as the following  (high-probability) upper and lower bound
\begin{align*}
    &
    \max_{i\in\I_k}\{\mu_{\GP_i}(z) -  \beta_{\GP_i}\sigma_{\GP_i}(z)\} 
    \leq g(z)
    \\
    \leq& \min_{i\in\I_k}\{\mu_{\GP_i}(z) + \beta_{\GP_i}\sigma_{\GP_i}(z)\}.
\end{align*}
Considering any function~$\bg$, it holds
\begin{align*}
    &
    \max_{i\in\I_k}\{\mu_{\GP_i}(z) -  \beta_{\GP_i}\sigma_{\GP_i}(z)\}
    -\bg(z)
    \\
    & \leq g(z)-\bg(z)
    \\
    &
    \leq \min_{i\in\I_k}\{\mu_{\GP_i}(z) + \beta_{\GP_i}\sigma_{\GP_i}(z)\} - \bg(z),
\end{align*}
which implies the following symmetric bound:
\begin{align*}
    &|g(x)-\bg(z)| 
    \nonumber
    \\
    \leq 
    &\max\left\{\min_{i\in\I_k}\{\mu_{\GP_i}(z) + \beta_{\GP_i}\sigma_{\GP_i}(z)\} - \bg(z)\right.,
    \nonumber
    \\
    &
    \phantom{\leq\max\left\{\right.}\left.
    \bg(z) - \max_{i\in\I_k}\{\mu_{\GP_i}(z) - \beta_{\GP_i}\sigma_{\GP_i}(z)\} \right\}.
\end{align*}
The monotonicity condition~\eqref{eq:set_inclusion_w} follows from the definition of~$\tilde{w}$~\eqref{eq:def_w_tilde_RAMPC} and the fact that~$\I_k\subseteq\I_{k+1}$.
Finally, consistency~\eqref{eq:g_bar_k_plus_one_g_bar} follows from~\eqref{eq:bar_g_as_weighted_sum} by setting~$\tilde{\lambda}_{i}=\lambda_{i}$ for all~$i\in \I_k$, and~$\tilde{\lambda}_i=0$ for all~\mbox{$i\in \I_{k+1}\backslash \I_k$}.~$\hfill \square$
}{}

\ifbool{arxiv}{
\subsection{Proof of Theorem~\ref{thm:RAMPC_tube}}\label{sec:proof_constraint_satisfaction_trajectory}
Given the bound on $\dot{V_\delta}$ derived in Proposition~\ref{prop:Lyapunov_function_dynamics}, this proof is analogous to ~\cite[Thm.~1]{sasfi2023RAMPC}.
Note that~\eqref{subeq:RMPC_result_1} holds at time~$t=0$ because of~\eqref{subeq:RMPC_assumption_3}. Suppose for contradiction that there exists a time~$\tau\geq 0$ for which $\Vdelta(x(t),z_\tau)\leq \delta_t$ holds for all~$t\leq\tau$ but is violated for~$t>\tau$. In this case,~\mbox{$\Vdelta(x(\tau),z_\tau)=\delta_{\tau}$} and~\mbox{$\dot{V}_{\delta}(x(\tau),z_\tau)>\dot{\delta}_\tau$}. Because~\mbox{$\Vdelta(x(t),z_t)=\delta_{t}$} at time~$t=\tau$, Lemma~\ref{lemma:constraint_satisfaction} in combination with Eq.~\eqref{subeq:RMPC_assumption_1} implies~\mbox{$(\gOpt, \gammaU)\in \Z, s\in[0,1]$}, with~$\gOpt$ being the geodesic of~$x(\tau)$,~$z_\tau$. Therefore, we can invoke Proposition~\ref{prop:Lyapunov_function_dynamics}, which bounds~$\dot{\Vdelta}(x(t),z_t)$ by the function~$f_\delta$~\eqref{subeq:RMPC_assumption_4_part2} (see Appendix~\ref{sec:preparatory_results}). Thus, we have
\begin{align*}  
    \dot{\Vdelta}(x(\tau),z_\tau)
    &
    \leq
    f_\delta\left(z_\tau,\Vdelta(x(\tau),z_\tau), w\right)
    \\
    &
    =
    f_\delta\left(z_\tau,\delta_\tau,w\right)
    \\
    &
     \stackrel{\mathmakebox[\widthof{=}]{\eqref{subeq:RMPC_assumption_4_part2}}}{=}
    \quad\dot{\delta}_\tau,
\end{align*}
which yields a contradiction with~\mbox{$\dot{\Vdelta}(x(\tau),z_\tau)>\dot{\delta}_\tau$}. Therefore,~\eqref{subeq:RMPC_result_1} holds for all~$t\geq0$. Lastly, Eqs.~\eqref{subeq:RMPC_result_1},~\eqref{subeq:RMPC_assumption_1} and Lemma~\ref{lemma:constraint_satisfaction} in Appendix~\ref{sec:preparatory_results} imply constraint satisfaction~\eqref{subeq:RMPC_result_2}.~$\hfill \square$
}{}
\ifbool{arxiv}{
\subsection{Proof of Theorem~\ref{thm:RAMPC_feasibility_convergence}}\label{sec:proof_theoretical_properties}
Given the robust prediction established in Theorem~\ref{thm:RAMPC_tube} and the monotonicity from Proposition~\ref{prop:bound_g_g_bar}, this proof follows the arguments in~\cite[Thm.~12]{sasfi2023RAMPC}.\\
\textbf{Part I [Recursive feasibility]:} Assume that the optimization problem~\eqref{eq:RAMPC_optimization_problem} is feasible at some time~$t_k$ for the state $x(t_k)$ and GP selection~$\I_k$,\mbox{$k\in \N_{\geq 0}$}. In the following, we construct a feasible candidate solution for time $t_{k+1}=t_k+T_{\mathrm{s}}$. 
We define 
\begin{alignat*}{2} 
    &v_{\tau|t_k}^\star :=\uref(\bg^\star_{T_{\mathrm{f}}|t_k}),\quad &&\tau\in[T_\mathrm{f},T_\mathrm{f}+T_\mathrm{s}),\\
    &\lambda^\star_{\tau+T_\mathrm{s}|t_{k},i}:=\lambda^\star_{T_\mathrm{f}|t_{k},i},\quad &&\tau\in[T_\mathrm{f},T_\mathrm{f}+T_\mathrm{s}),
\end{alignat*}
with~$z_{\tau|t_k}^\star$ and~$\delta_{\tau|t_k}^\star$ according to~\eqref{subeq:RMPC_formulation_constraint_1} and~\eqref{subeq:RMPC_formulation_constraint_2} respectively. 
At time~$t_{k+1}$, with GP selection~$\I_{k+1}$ and measured state $x(t_{k+1})$, we consider the candidate solution
\begin{subequations}\label{eq:candidate_solution_RAMPC}
\begin{align}
    v_{\tau|t_{k+1}} 
    &
    = v_{\tau+T_\mathrm{s}|t_k}^\star\quad \tau\in[0,T_\mathrm{f}],
    \\
    z_{0|t_{k+1}}
    &
    = z_{T_\mathrm{s}|t_k}^\star,
    \\
    \delta_{0|t_{k+1}}
    &
    = \Vdelta(x(t_{k+1}),z_{T_\mathrm{s}|t_k}^\star),
\end{align}
\end{subequations}
together with the trajectories~$z_{\tau|t_{k+1}}$,~$\delta_{\tau|t_{k+1}}$,~$\tau\in[0,T_\mathrm{f}]$ according to~\eqref{subeq:RMPC_formulation_constraint_1} and~\eqref{subeq:RMPC_formulation_constraint_2} respectively. In addition, we consider
\begin{subequations} \label{eq:candidate_solution_lambda}
\begin{alignat}{2}
    &\lambda_{\tau|t_{k+1},i} = \lambda^\star_{\tau+T_\mathrm{s}|t_{k},i},\; &&\forall \tau\in[0,T_\mathrm{f}], i\in \I_k \cap\I_{k+1},
    \\
    &\lambda_{\tau|t_{k+1},i} =0,\quad&&\forall \tau\in[0,T_\mathrm{f}], i\in \I_{k+1}\backslash\I_k,
\end{alignat}
\end{subequations}
which ensures the consistency property~\eqref{eq:g_bar_k_plus_one_g_bar} of Proposition~\ref{prop:bound_g_g_bar}. Note that the indices~$i\in\I_{k}\backslash\I_{k+1}$ eliminated by Algorithm~\ref{alg:GP_selection} corresponds to cases where~$\lambda_{\tau|t_k,i}=0$ for all $\tau\in[0,T_\mathrm{f}]$, so that they do not contribute to the expression of~$\bg$~\eqref{eq:bar_g_as_weighted_sum}, i.e., $\bg_{\tau|t_{k+1}}=\bg_{\tauts|t_{k}}^\star$.  Furthermore,  the monotonicity property~\eqref{eq:set_inclusion_w} ensures $w_{\tau|t_{k+1}}(\cdot)\leq w^\star_{\tau+T_{\mathrm{s}}|t_k}(\cdot)$.
The resulting nominal state trajectory satisfies
%
    $z_{\tau|t_{k+1}}=z_{\tauts|t_k}^\star$, $\tau\in[0,T_\mathrm{f}]$.
Next, we show that the new tube scaling $\delta$ satisfies
\begin{equation}\label{eq:candidate_solution_delta_inequality_RAMPC}
    \delta_{\tau|t_{k+1}}\leq \delta_{\tauts|t_k}^\star,\quad \tau\in[0,T_\mathrm{f}],
\end{equation}
i.e., the candidate prediction is contained in the previous prediction. 
First, Theorem~\ref{thm:RAMPC_tube} bounds the deviation between the nominal prediction and state trajectory as
\begin{equation}\label{eq:Vdelta_leq_delta_0_Ts_RAMPC}
    \Vdelta(x(t_k+\tau),z_{\tau|t_k}^\star)\leq \delta_{\tau|t_k}^\star,\quad \tau\in[0,T_\mathrm{s}].
\end{equation}
Hence, the initial value of the tube scaling satisfies
\begin{equation*}
    \delta_{0|t_{k+1}} = \Vdelta(x(t_{k+1}),z^\star_{T_\mathrm{s}|t_k})\leq \delta_{T_\mathrm{s}|t_k}^\star.
\end{equation*}
For contradiction, suppose that there exists a time~$\tauhat\in[0,T_\mathrm{f})$, such that condition~\eqref{eq:candidate_solution_delta_inequality_RAMPC} holds for all~$\tau \in [0,\tauhat]$, but is violated for~$\tau>\tauhat$, i.e.,~$\delta_{\tauhat|t_{k+1}}=\delta_{\tauhatTs|t_k}^\star$ and~$\dot{\delta}_{\tauhat|t_{k+1}}>\dot{\delta}_{\tauhatTs|t_k}^\star$. Using the monotonicity property~\eqref{eq:set_inclusion_w} of Proposition~\ref{prop:bound_g_g_bar}, the dynamics of the tube scaling~\eqref{subeq:RMPC_assumption_4_part2} yields
    {\allowdisplaybreaks
\begin{align*}
        \dot{\delta}_{\tauhat|t_{k+1}} 
        &
        = f_\delta(z_{\tauhatTs|t_k}^\star,\delta_{\tauhat|t_{k+1}},
        w_{\tauhat|t_{k+1}})
        \\
        &
        \stackrel{\mathmakebox[\widthof{=}]{\eqref{eq:set_inclusion_w}}}{\leq}f_\delta(z_{\tauhatTs|t_k}^\star,\delta_{\tauhat|t_{k+1}}, w^\star_{\tauhatTs|t_k})
        \\
        &
        =f_\delta(z_{\tauhatTs|t_k}^\star,\delta^\star_{\tauhatTs|t_{k}},
        w^\star_{\tauhatTs|t_k})
        \\
        &
        =\dot{\delta}_{\tauhatTs|t_k}^\star,
        \end{align*}}
which is a contradiction. Hence,~\eqref{eq:candidate_solution_delta_inequality_RAMPC} holds.

 We now leverage the terminal set conditions (Asm.~\ref{assumption:terminal_set}) to show feasibility of the candidate solution  for~$\tau\in[T_\mathrm{f}-T_\mathrm{s},T_\mathrm{f}]$.
 The candidate solution satisfies the terminal set constraint~\eqref{subeq:RMPC_formulation_constraint_4} due to the positive invariance~\eqref{eq:terminal_set_PI}, the monotonicity property~\eqref{eq:terminal_set_monotonicity} of the terminal set~$\Xterminal$,~\eqref{eq:candidate_solution_delta_inequality_RAMPC} and $w_{\tau|t_{k+1}}(\cdot)\leq w^\star_{\tau+T_{\mathrm{s}}|t_k}(\cdot)$. The constraint satisfaction condition~\eqref{eq:terminal_set_constraint_satisfaction} of the terminal set coupled with~\eqref{eq:candidate_solution_delta_inequality_RAMPC} also implies satisfaction of the tightened constraints~\eqref{subeq:RMPC_formulation_constraint_3} by the candidate solution for~$\tau\in[T_\mathrm{f}-T_\mathrm{s},T_\mathrm{f}]$.

For~$\tau\in[0,T_\mathrm{f}-T_\mathrm{s}]$, the tightened constraints~\eqref{subeq:RMPC_formulation_constraint_3} are also satisfied by the candidate solution, as
\begin{align*}
    & h_j(z_{\tau|t_{k+1}},v_{\tau|t_{k+1}})+c_j\delta_{\tau|t_{k+1}}
    \\
    &
    \stackrel{\mathmakebox[\widthof{=}]{\eqref{eq:candidate_solution_delta_inequality_RAMPC}}}{\leq}
    h_j(z^\star_{\tau+Ts|t_{k}},v^\star_{\tau+T_\mathrm{s}|t_{k}})+c_j\delta^\star_{\tau+T_\mathrm{s}|t_{k}}
    \stackrel{\mathmakebox[\widthof{=}]{\eqref{subeq:RMPC_formulation_constraint_3}}}{\leq}
    0.
\end{align*}
}{}
\ifbool{arxiv}{
\textbf{Part II [Constraint satisfaction]:}
Given recursive feasibility of the MPC scheme, satisfaction of the state and input constraints~\eqref{eq:state_input_constraints} for
~$t\in[t_k,t_k+T_\mathrm{s}]$, $k\in\mathbb{N}$, directly follows with Theorem~\ref{thm:RAMPC_tube} due to the tightened constraints~\eqref{subeq:RMPC_formulation_constraint_3}.
}{}
\ifbool{arxiv}{
\\\textbf{Part III [Convergence]:} 
Let us denote the optimal cost to Problem~\eqref{eq:RMPC_optimization_problem} at time $k$ by $V^\star(x(t_k),\I_k)$.
The candidate solution~\eqref{eq:candidate_solution_RAMPC}-\eqref{eq:candidate_solution_lambda} provides an upper bound on the optimal cost at time~$t_{k+1}$:
{\allowdisplaybreaks\begin{align*}
    &V^\star(x(t_{k+1},\I_{k+1}))
    \\
    \leq& \> \int_0^{T_\mathrm{f}}\ell (z_{\tauts|t_{k}}^\star,v_{\tauts|t_{k}}^\star,\bg_{\tauts|t_{k}}^\star)\dtau\\
    &+ \ell_\mathrm{f}(z_{T_\mathrm{f}+T_\mathrm{s}|t_{k}}^\star,\bg_{T_\mathrm{f}+T_\mathrm{s}|t_{k}}^\star)
    \\
    =& \> V^\star(x(t_{k},\I_k)) -\int_0^{T_\mathrm{s}}\ell (z_{\tau|t_{k}}^\star,v_{\tau|t_{k}}^\star,\bg_{\tau|t_{k}}^\star)\dtau \\
    &\ell_\mathrm{f}(z_{T_\mathrm{f}+T_\mathrm{s}|t_{k}}^\star,\bg_{T_\mathrm{f}+T_\mathrm{s}|t_{k}}^\star)-\ell_\mathrm{f}(z_{T_\mathrm{f}|t_{k}}^\star,\bg_{T_{\mathrm{f}}|t_{k}}^\star)
    \\
    &\> +\int_{T_\mathrm{f}}^{T_\mathrm{f}+T_\mathrm{s}}\ell (z_{\tau|t_{k}}^\star,v_{\tau|t_{k}}^\star,\bg_{\tau|t_{k}}^\star)\dtau
    \\
    \stackrel{\mathmakebox[\widthof{=}]{\eqref{eq:terminal_set_local_CLF}}}{\leq}& \>
    V^\star(x(t_{k}),\I_k) -\int_0^{T_\mathrm{s}}\ell (z_{\tau|t_{k}}^\star,v_{\tau|t_{k}}^\star,\bg_{\tau|t_{k}}^\star)\dtau. 
\end{align*}}
Using the previous inequality in a telescopic sum yields
\begin{align}\label{eq:convergence_recursive_inequality_RAMPC}
     &V^\star(x_0,\I_0) -\limsup_{k\rightarrow \infty} V^\star(x(t_k),\I_k)
     \nonumber
     \\
    &\geq
    \sum_{k=0}^\infty \int_0^{T_\mathrm{s}} \ell(z_{\tau|t_k}^\star,v_{\tau|t_k}^\star,\bg_{\tau|t_k})\dtau.
\end{align}
As the optimization problem~\eqref{eq:RAMPC_optimization_problem} is subject to compact constraints and its cost~$\ell$ is continuous, the term~$\limsup_{k\rightarrow \infty} V_{T_\mathrm{f}}^\star(x(t_k))$ is uniformly bounded. Hence, the right-hand side of~\eqref{eq:convergence_recursive_inequality_RAMPC} is bounded, and 
%
\begin{equation}\label{eq:lim_int_l_RAMPC}
 \lim_{k\rightarrow \infty} \int_0^{T_\mathrm{s}} \ell(z_{\tau|t_k}^\star,v_{\tau|t_k}^\star,\bg^\star_{\tau|t_k})\dtau = 0.
\end{equation}
To show that~\eqref{eq:lim_int_l_RAMPC} leads to convergence, we note that~$\ell(z_{\tau|t_k}^\star,v_{\tau|t_k}^\star,\bg_{\tau|t_k})$ is uniformly continuous. Indeed,~$z_{\tau|t_k}^\star$ is uniformly continuous in~$\tau$, given that~$\bar{f}$ is Lipschitz and~$(z_{\tau|t_k}^\star,v_{\tau|t_k}^\star)$ is subject to compact constraints~\eqref{subeq:RMPC_formulation_constraint_3}, as well as~$v^\star_{\tau|t_k}, \bg^\star_{\tau|t_k}$ being constant on~$\tau\in[0,T_\mathrm{s})$. Since~$\ell$ is uniformly continuous and non-negative,~\eqref{eq:lim_int_l_RAMPC} yields
\begin{equation*}
 \lim_{k\rightarrow \infty}  \ell(z_{\tau|t_k}^\star,v_{\tau|t_k}^\star,\bg^\star_{\tau|t_k}) = 0, \quad\tau\in[0,T_\mathrm{s}).
\end{equation*}
In addition, and because~$\ell$ is positive definite as expressed by the quadratic cost~\eqref{eq:def_quadratic_stage_cost}, it holds 
\begin{equation*}
\lim_{k\rightarrow \infty}\norm{(z^\star_{\tau|t_k},v^\star_{\tau|t_k})-(x_\mathrm{ref}(\bg^\star_{\tau|t_k}),u_\mathrm{ref}(\bg^\star_{\tau|t_k}))}=0, 
\end{equation*}~for all $\tau\in[0,T_\mathrm{s})$ $\hfill \square$
}{}

\ifbool{arxiv}{
\subsection{\RMPC{} as a special case}\label{sec:proof_RMPC_special_case}

The \RMPC{} formulation is a special case of the more general \RAMPC{} scheme, where the model selection~$\I_k$ contains a single GP model trained on the offline data, and the corresponding weight~$\lambda_{t,0}=1$ for all~$t\geq0$. The proofs remain valid for this special case.
}{}

\ifbool{arxiv}{
\section{Adaptation for vector-valued function~$g$}\label{sec:vector_valued_g}

In the following, we adapt our theoretical results to the case where the unknown function~$g$ is a vector-valued function, i.e., $g(x)\in\mathbb{R}^l$, $G\in\mathbb{R}^{n\times l}$. We model such cases using independent GPs for each of the (output) dimensions of~$g$. 
For ease of notation, we consider the GP mean $\mu_{N,i}$ and standard deviation $\sigma_{N,i}$, $i\in\N_{[1,l]}$ using the $N$ offline measurements, while additional online measurements are handled analogous to Section~\ref{sec:RAMPC_framework}.
This allows to modify Proposition~\ref{prop:Lyapunov_function_dynamics}, as follows.
%
\begin{proposition}\label{prop:Lyapunov_function_dynamics_multi_output_g}
    Suppose Assumption~\ref{assumption:CCM} holds. In addition, suppose that Assumptions~\ref{assumption:measurement_noise_sequence_g_RKHS} holds for each dimension~$i\in\N_{[1,l]}$ of the unknown function~$g(x)\in \R^l$ and the noise $\epsilon\in\R^l$ independently. 
    Then, 
    \begin{align}
    \label{eq:uncertain_bound_multi}
    &\mathbb{P} \big\{ |g_i(x)-\mu_{N,i}(x)| \leq w_i(x)
    \nonumber,
    \\
    &\quad\quad\forall x\in \mathbb{Z}_x, N\in\N_{\geq 0},i\in\N_{[1,l]} \big\}\nonumber\\
     & \geq \prod_{i=1}^l (1-p_i)=:1-p
    \end{align}
    with $w_i(x)=\beta_{N,i}\sigma_{N,i}(x)$ and $\beta_{N,i}$ according to~\eqref{eq:uncertainty_bound_fiedler} with probability $p_i$ and $\|g_i\|_{\Hilbert_k}\leq B_{g,i}$ for each dimension~$i$.
    Then, for all~$x,z\in \Z_x$ such that~$(\gOpt,\gammaU)\in\Z$ for~$s\in[0,1]$,  the following bound holds with a probability of at least~$1-p$:
    \begin{subequations}\label{eq:prop_Lyapunov_function_dynamics_multi_output_g}
        \begin{align}
            \dot{\Vdelta} (x,z)
            =
            &
            \pder{\Vdelta}{x}\Bigr|_{(x,z)} \xd + \pder{\Vdelta}{z}\Bigr|_{(x,z)} \zd
            \\
            \leq
            &
            -\left(\rho - L_G \right)\Vdelta(x,z) 
            \nonumber
            \\
            &+ \sum_{i=1}^l G_{M,i} w_i(z) + E_M,
        \end{align}
    \end{subequations}
    where~$L_G$ and~$E_M$ are defined in~\eqref{eq:def_offline_constants}, $\dot{x}=f_w(x,u,g,d)$, $\dot{z}=\bar{f}(z,v,\bar{g})$, $\bar{g}(z)=[\mu_{N,1}(z),\dots, \mu_{N,l}(z)]^\top$,  and
    %
    \begin{equation}
        \label{subeq:def_G_M_multi_output_g}
            G_{M,i}
            :=
            \max_{x\in \Z_x}\left\{ \norm{[\Mhalfarg{x}G]_{:,i}} \right\},
    \end{equation}
    with~$[\Mhalfarg{x}G]_{:,i}$ denoting the $i$-th column of~$\Mhalfarg{x}G$.
\end{proposition}

\begin{pf}
 The high-probability error bound~\eqref{eq:uncertain_bound_multi} follows directly from Lemma~\ref{lemma:GP_estimate_uncertainty_bound}, given that each dimension is treated independently and the noise is independent across dimensions. 
    The bound~\eqref{eq:prop_Lyapunov_function_dynamics_multi_output_g} is derived analogously to Proposition~\ref{prop:Lyapunov_function_dynamics}, using a summation over the uncertainty bounds for each dimension~$i$. In particular, the derivation  between inequality~\eqref{subeq:proof_tube_dynamics_last_part_3} and~\eqref{subeq:proof_tube_dynamics_last_part_4} is modified as follows:
    \begin{subequations}
        \begin{align*}
            &
            \max_{s\in[0,1]} \norm{\Mgammahalf{s} \gammaW} 
            \\
            \leq
            &
            \max_{s\in [0,1]}\left\{ \norm{ \Mgammahalf{s}G(g(x)-g(z))} \right.
            \nonumber
            \\
            & + \norm{\Mgammahalf{s}G(g(z)-\mu_N(z))}
            \nonumber
            \\
            & \left. + \norm{{\Mgammahalf{s}E(\gOpt)d}} \right\}
            \\
            \leq
            &
            \max_{s\in [0,1]}\left\{ \norm{ \Mgammahalf{s}G(g(x)-g(z))}\right\}
            \nonumber
            \\
            & +  \sum_{i=1}^l\max_{s\in [0,1]}\left\{\norm{[\Mgammahalf{s}G]_{:,i}} \right.
            \nonumber
            \\
            &\left.|g_i(z)-\mu_{N,i}(z)|\right\}
            \nonumber
            \\
            & + \max_{s\in [0,1]}\left\{\norm{{\Mgammahalf{s}E(\gOpt)d}} \right\}
            \\
            \stackrel{\mathmakebox[\widthof{=}]{\substack{\eqref{eq:def_offline_constants}\eqref{subeq:def_G_M_multi_output_g}}}}{\leq}
            &
            \quad\;\; L_G \Vdelta(x,z) + \sum_{i=1}^l G_{M,i} w_i(z) + E_M.\quad  \hfill \square
        \end{align*}
    \end{subequations}
\end{pf}
 Thanks to Proposition~\ref{prop:Lyapunov_function_dynamics_multi_output_g}, the  design and the theoretical guarantees derived in Sections~\ref{sec:RMPC_framework} and~\ref{sec:RAMPC_framework} follow naturally with minor differences. 
 We note that for the RAMPC scheme, each uncertainty bound~$w_i$ needs to be computed separately according to~\eqref{eq:prop_bound_g_g_bar}. 
    
}{}

\ifbool{arxiv}{
\section{Supplementary material}\label{sec:supplementary_material}
\subsection{Selection of probability bounds}\label{sec:probability_bounds_table}
A selection of different bounds~$w$ under different noise assumptions is given in the following table. 
\begin{table}[htb]
\begin{center}
\small{
\begin{tabular}{@{}cllc@{}}
\toprule
Ref.                            & Noise assumptions& Authors & Year                            \\ \midrule
\cite{molodchyk2025towards}         & Indep. Gaussian noise & Molodchyk et al.  & 2025        \\
\cite{srinivas2012information}      & Gaussian noise        & Srinivas et al.   & 2012                             \\
\cite{abbasi2013online}             & Cond.~$R$-sub-Gaussian & Abbasi-Yadkori   & 2013      \\
\cite{chowdhury2017kernelized}      & Cond.~$R$-sub-Gaussian& Chowdhury et al.  & 2017    \\
\cite{fiedler2021practical}         & Cond.~$R$-sub-Gaussian& Fiedler et al.    & 2021          \\
\cite{maddalena2021deterministic}   & Point-wise bounded noise& Maddalena et al.& 2021         \\ 
\cite{lahr2025optimal}           & Energy-bounded noise    & Lahr et al.      &2025                                 \\ \bottomrule
\end{tabular}}

\end{center}
\end{table}\\
}{}

\end{document}